\documentclass{article}
\usepackage[utf8]{inputenc}
\usepackage{amsmath}
\usepackage{amsfonts}
\usepackage{upgreek}
\usepackage{bm}
\usepackage{xcolor}
\usepackage{comment}
\usepackage{graphicx}
\usepackage{biblatex}
\DeclareMathOperator{\diag}{diag}

\renewcommand{\det}[1]{\left|#1\right|}

\renewcommand{\vec}[1]{\boldsymbol{\mathbf{#1}}}
\newcommand{\mat}[1]{\boldsymbol{\mathbf{#1}}}

\title{The Global Structure of Codimension-2 Local Bifurcations in Continuous-Time Recurrent Neural Networks}
\author{Randall D. Beer\\ \small Cognitive Science Program\\ \small Luddy School of Informatics, Computing, and Engineering\\ \small Indiana University\\ \small Bloomington, Indiana\\ \small rdbeer@iu.edu}

\begin{document}
\maketitle

\begin{abstract}
If we are ever to move beyond the study of isolated special cases in theoretical neuroscience, we need
to develop more general theories of neural circuits over a given neural model. The present paper considers this challenge in the context of 
continuous-time recurrent
neural networks (CTRNNs), a simple but dynamically-universal model that has been widely utilized in both 
computational neuroscience and neural networks. Here we extend previous work on the parameter space 
structure of codimension-1 local bifurcations in CTRNNs to include codimension-2 local bifurcation manifolds.
Specifically, we derive the necessary conditions for all generic
local codimension-2 bifurcations for general CTRNNs, specialize these conditions to circuits
containing from one to four neurons, illustrate in full detail the application of these conditions to 
example circuits, derive closed-form expressions for these bifurcation manifolds where possible, and
demonstrate how this analysis allows us to find and trace several global codimension-1 bifurcation manifolds
that originate from the codimension-2 bifurcations.
\end{abstract}

\section{Introduction}

Continuous-time recurrent neural networks (CTRNNs) are neural models of the form

\begin{equation}
\vec{\uptau}\dot{\vec{y}} = -\vec{y}+\mat{W}\psi \left(\vec{y}+\vec{\uptheta}\right) + \vec{I}
\end{equation}

\noindent
where $\vec{y}$, $\vec{\uptau}$, $\vec{\uptheta}$, and $\vec{I}$ are length $n$ vectors
describing the network state, time constants, 
biases and external inputs, respectively, $\psi(\cdot)$ is a smooth, monotonically-increasing, bounded
sigmoidal function, and $\mat{W}$ is an $n\times n$ weight matrix describing the coupling between neurons.
Many different sigmoidal functions have been described in the literature. When a specific choice
must be made for concreteness, we will use $\sigma(x) \equiv 1/(1+e^{-x})$, referring to the
resulting model as $\sigma$-CTRNNs.

CTRNNs can also written as

\begin{equation*}
\vec{\uptau}\dot{\vec{y}} = -\vec{y} + \psi \left(\mat{W} \vec{y}+\vec{\uptheta} + \vec{I}\right) 
\end{equation*}

\noindent
but the two formulations are completely equivalent under the substitution
$\vec{y} \mapsto \psi(\mat{W}\vec{y} + \mat{\theta} + \vec{I})$.
This second formulation makes clear that it is only the net
input $\vec{I}+\vec{\uptheta}$ that matters to the dynamics. Accordingly, we will 
absorb the external input $\vec{I}$ into our definition of $\vec{\uptheta}$. 
In addition, we will henceforth assume that all time constants are unity.
Although all derivations in this paper can be carried
out for arbitrary time constants, the final expressions often become considerably
more unwieldy.

CTRNNs are arguably the simplest continuous-time model that captures the nonlinear dynamics and 
recurrent connectivity of nervous systems. As such, they support a variety of neurobiological
applications and interpretations \cite{barak,sussillo}, including, most commonly, as models of the mean 
firing rate properties of spiking nerve cells 
\cite{ermentrout}, as models of graded synaptic transmission in nonspiking nerve cells \cite{izquierdo10, olivares}, 
and as models of interacting homogeneous populations of nerve cells \cite{wilson,fasoli}. 
Recurrent neural network models such as CTRNNs have also been employed in the application of artificial neural networks to problems
involving associative memory \cite{hopfield}, combinatorial optimization \cite{hopfieldtank,smithmiles}, the generation
or processing of temporal sequences \cite{yu}, evolutionary robotics \cite{beer92,floreano} and reservoir
computing \cite{maass,jaeger}. Indeed, since CTRNNs 
are known to be universal dynamics approximators \cite{funahashi,kimura,chow}, they can also be interpreted as
simply convenient building blocks out of which any desired dynamics can be constructed. 
CTRNNs and closely-related models have been the subject of extensive dynamical analysis \cite{grossberg,wilson,cowan,cohen,hopfield,sompolinsky,hirsch,das,borisyuk,blum,zhaojue,hoppensteadt,tino,pasemann,haschke,fasoli,cervantes-ojeda}.

The work described in this paper represents the latest step in a long-term research program aimed at systematically
developing a general theory of neural circuits over the CTRNN neural model through characterizing its
parameter space structure as exhaustively as possible \cite{beer95,beer06,beer10}. Although CTRNNs are quite
simple, progress on this ``toy model" can not only directly benefit applications where this specific 
neural model is already used, but may also contribute to building intuition and tools for 
understanding the parameter space structure of more biophysically realistic models.
As Gao and Ganguli \cite{gao} explain, ``An even higher level of
understanding is achieved when we develop not just a single model that explains a data set,
but rather understand the space of all possible models consistent with the data. Such an 
understanding can place existing biological systems within their evolutionary context, 
leading to insights about why they are structured the way they are, and can reveal general 
principles that transcend any particular model."

Bifurcations in nonlinear dynamical systems can be stratified by \emph{codimension}. 
In general, the term ``codimension" refers to the dimension of some subspace relative to the ambient space that contains it.
Thus, for example, a surface embedded in a 3-dimensional space has codimension 1, 
whereas a point in the same space has codimension 3. 
Similarly, the codimension of a bifurcation in some system is the dimension of the subset 
of parameter values for which the system exhibits that bifurcation relative to the dimension 
of its entire parameter space.
More formally, the codimension of a bifurcation is the minimum number of parameters appearing 
in its universal unfolding \cite[pp. 392--415]{wiggins}.
Roughly speaking, the codimension of a bifurcation tells us how typical it is, 
with higher codimension bifurcations becoming increasingly rare. Normally, only 
dynamical systems exhibiting low-codimension
behavior are considered to be good mathematical models of real-world phenomena, since 
they are the only ones whose behavior is robust to the inevitable parametric uncertainties.

If models exhibiting higher codimension bifurcations are to be avoided in applications, 
what is the point in studying them? There are three main reasons. First, experimental uncertainties
induce distributions over a model's parameters which, in high-dimensional parameter spaces, 
can be large enough to generically overlap higher-codimension bifurcations. 
Thus, even if we prefer to avoid such structurally-unstable models, we have may no choice but to 
consider their consequences in applications.
Second, symmetries -- which are common in the natural world -- can 
lower the codimension of a bifurcation, making normally atypical bifurcations typical. For example,
pitchfork bifurcations become codimension-1 in CTRNNs when the center-crossing symmetry \cite{mathayomchan}
is applied.
Third, and most importantly, higher codimension bifurcations serve as \emph{organizing centers} that anchor
lower-codimension bifurcations in much the same way that cities anchor the road network of a country. 
Knowing the locations and types of such organizing centers thus tells us something about the arrangement of
lower-codimension bifurcations. This is especially important for identifying the locations and layout
of global bifurcations, since organizing centers can be the only local way in which to find such behavior.

In this paper, we extend previous work on characterizing the global structure of local
codimension-1 bifurcation manifolds in CTRNNs to encompass the global structure of
local codimension-2 bifurcations as well. The next section reviews some basic concepts and notation
that is required for the remainder of the paper.
Next, we give the necessary conditions for each of the two generic local codimension-1 
and the five generic local codimension-2 bifurcations in turn and then specialize those conditions
to CTRNNs. We then work out these general conditions in detail for 1-, 2-, 3- and 4-neuron CTRNNs. 
In each case, we first derive the corresponding conditions for general CTRNNs of that size and then
specialize those condition to $\sigma$-CTRNNs. We also work through a specific example for each size 
in order to illustrate the calculations. The paper ends with a brief discussion of the current status
and future directions of this line of research.

\section{Preliminaries}

Following Haschke and Steil \cite{haschke}, our bifurcation analysis will be performed in the
space of activation function derivatives $\vec{\uppsi}' \equiv (\psi'_1, \ldots, \psi'_n)$, 
where each $\psi'_i \equiv \psi'(y_i^*+\theta_i)$ is evaluated at an equilibrium
point $\vec{y}^* \equiv (y_1^*,\ldots,y_n^*)$. The advantage of working in $\vec{\uppsi}'$ is
that the conditions defining various bifurcations are more directly and simply expressed.
As a final step, results obtained in $\vec{\uppsi}'$ can be mapped into net input space 
$\vec{\uptheta}$ using

\begin{equation}
\vec{\uptheta} = \psi'^{-1}(\vec{\uppsi}') - \mathbf{W} \psi(\psi'^{-1}(\vec{\uppsi}'))
\end{equation}

\noindent
where we think of the bifurcation manifolds in net input space as being parameterized by
$\mat{W}$.

For $\sigma$-CTRNNs this becomes

\begin{equation}
\label{eq:SigmaTheta}
\vec{\uptheta} = \vec{\Uptheta}(\vec{\upsigma}') \equiv \sigma'^{-1}(\vec{\upsigma}') - \mathbf{W} \sigma(\sigma'^{-1}(\vec{\upsigma}'))
\end{equation}

\noindent
where

\begin{equation}
\label{eq:DSigmaInverse}
\sigma'^{-1}(\sigma'_i) = \ln \frac{1 \pm \sqrt{1 -  4 \sigma'_i} - 2 \sigma'_i}{2\sigma'_i}
\end{equation}

\noindent
Note that $\sigma'^{-1}(\cdot)$ is 2-valued. Since each component of $\vec{\uptheta}$ can
come from either branch, each bifurcation manifold in $\vec{\upsigma'}$-space can give rise to up to
$2^n$ bifurcation manifolds in $\vec{\uptheta}$-space and thus $\vec{\Uptheta}(\vec{\upsigma}')$ is
in general $2^n$-valued.
The domain of $\sigma'^{-1}(\cdot)$ is also restricted as $0<\sigma'\le 1/4$.

Finally, a convenient property of $\sigma(\cdot)$
that we will make use of is that its higher derivatives can be written in terms of $\sigma'$:

\begin{equation}
\label{eq:PsiPrimeSubst}
\sigma'' = \pm \sigma' \sqrt{1 - 4 \sigma'}
\end{equation}

\begin{equation}
\label{eq:PsiPrimePrimeSubst}
\sigma''' = \sigma' - 6 {\sigma'}^2
\end{equation}

Our formulation of local bifurcation theory will closely follow Kuznetsov \cite{kuznetsov}.
In the neighborhood of an equilibrium point of $\dot{\vec{x}} = \vec{f}(\vec{x})$,
$\vec{f}(\vec{x})$ can be written as a Taylor expansion

\begin{equation*}
    \vec{f}(\vec{x}) = \mat{A}\vec{x}+\frac{1}{2}\vec{B}(\vec{x},\vec{x})+\frac{1}{6}\vec{C}(\vec{x},\vec{x},\vec{x})+\cdots
\end{equation*}

\noindent
where $\mat{A}$ is the $n \times n$ Jacobian matrix defined as

\begin{equation*}
\mat{A} \equiv \vec{\partial}_{\vec{x}} \vec{f}
\end{equation*}

\noindent
and $\mat{B}$ and $\mat{C}$ are multilinear vector functions whose components are given by

\begin{equation*}
B_i(\vec{u},\vec{v}) 
\equiv 
\sum_{j,k=1}^n \frac{\partial^2 f_i(\vec{\upxi})}{\partial \xi_j \partial \xi_k } u_j v_k
\end{equation*}

\begin{equation*}
C_i\left(\vec{u},\vec{v},\vec{w}\right)
\equiv
\sum_{y,k,l=1}^{n}\frac{\partial^3 f_i\left(\vec{\upxi}\right)}{\partial\xi_j\partial\xi_k\partial\xi_l}u_j v_k w_l
\end{equation*}

For CTRNNs, these expressions become

\begin{equation}
\mat{A} = \mat{W}\diag(\vec{\uppsi}') - \mat{1}
\end{equation}

\begin{equation}
\vec{B}(\vec{u},\vec{v})  = \mat{W}\vec{\uppsi}''\vec{u} \vec{v}
\end{equation}

\begin{equation}
\vec{C}\left(\vec{u},\vec{v},\vec{w}\right) =\mat{W}\vec{\uppsi}'''\vec{u}\vec{v}\vec{w}
\end{equation}

\noindent
where $\mat{1}$ denotes the $n \times n$ identity matrix, $\vec{\uppsi}'$, $\vec{\uppsi}''$ 
and $\vec{\uppsi}'''$ denote vectors of the first, second and third derivatives, 
respectively, of the function $\psi$ evaluated at an equilibrium point, and vector adjacency
denotes the Hadamard (element-wise) product of vectors.

We will think of bifurcation manifolds as (typically infinite) sets of points and denote them using
script capital letters. Thus, if $\vec{f}(\vec{x};\vec{p}) = 0$ is the defining condition for a
particular bifurcation manifold $\mathcal{B}$, we would denote it as $\mathcal{B}: \{\vec{f}=0\}$,
where $\{\vec{f}=0\}$ is an abbreviation for $\{\vec{x}\in\mathbb{R}^n, \vec{p}\in\mathbb{R}^m|\vec{f}(\vec{x},\vec{p})=0\}$.

We denote by $\mat{M}_{ij}$ the $(n-1)\times (n-1)$ submatrix formed by removing the $i$th row and the $j$th column from the $n\times n$ matrix $\mat{M}$.

The discriminant $\Delta_n$ of the characteristic polynomial of an $n\times n$ matrix $\mat{A}$ is defined for $n=2, 3, 4$ as

\begin{equation*}
\begin{split}
\Delta_2 \equiv~& \tau^2 - 4\det{\mat{A}}\\
\Delta_3 \equiv~& -27 \det{\mat{A}}^2 + \mu_2^2 (\tau^2 - 4 \mu_2) + 2 \det{\mat{A}} \tau (9 \mu_2 - 2 \tau^2)\\
\Delta_4 \equiv~& 256 \det{\mat{A}}^3 - 
\det{\mat{A}}^2 (128 \mu_2^2 + 192 \mu_3 \tau - 144 \mu_2 \tau^2 + 27 \tau^4) + \\
&~2 \det{\mat{A}} (8 \mu_2^4 - 40 \mu_2^2 \mu_3 \tau - 2 \mu_2^3 \tau^2 - 3 \mu_3^2 \tau^2 + 9 \mu_2 \mu_3 (8 \mu_3 + \tau^3)) +\\
&\mu_3^2 (-4 \mu_2^3 + 18 \mu_2 \mu_3 \tau + \mu_2^2 \tau^2 - \mu_3 (27 \mu_3 + 4 \tau^3))
\end{split}
\end{equation*}

\noindent
where $\tau$ and $\mu_k$ denotes the trace of $\mat{A}$ and the sum of the size-$k$ principal minors of $\mat{A}$, respectively. Note that $\tau\equiv\mu_1$.

Finally, the inner product of two complex vectors is defined as

\begin{equation*}
\langle\vec{u}, \vec{v}\rangle \equiv \vec{\overline{u}}\cdot \vec{v}
\end{equation*}

\noindent
where an overbar indicates complex conjugation.

\section{Local Codimension-1 Bifurcation Conditions}

Although our focus in this paper is on generic codimension-2 local bifurcations, we will
need to make use of the defining conditions of codimension-1 local bifurcations.
Generically, there are two types of codimension-1 local bifurcations, the \emph{saddle-node}
(aka fold or limit point) bifurcation and the \emph{Hopf} (aka Andronov-Hopf) bifurcation.
We briefly review each in turn. The local codimension-1 parameter space structure of CTRNNs
has been analyzed previously \cite{beer95,beer06,beer10}.

\subsection{Saddle-Node Bifurcation}

A saddle-node bifurcation occurs when a single real eigenvalue of the Jacobian matrix $\mat{A}$ passes through 0.
At a generic saddle-node bifurcation, either a single equilibrium point splits into a pair of equilibria, or a
pair of equilibria coalesce into one. A necessary condition for this to occur is

\begin{equation*}
\det{\mat{A}} = 0
\end{equation*}

\noindent
with additional nondegeneracy and transversality conditions required to ensure sufficiency.

For general CTRNNs, the saddle-node bifurcation condition becomes

\begin{equation}
\label{eq:SN}
\mathcal{S}:~\{\det{\mat{W}\diag(\vec{\uppsi'}) - \mat{1}} = 0\}
\end{equation}

\subsection{Hopf Bifurcation}

A Hopf bifurcation occurs when the real parts of a single pair of complex conjugate eigenvalues
of the Jacobian matrix $\mat{A}$ pass through 0. At a generic Hopf bifurcation, an equilibrium point
changes stability, either giving rise to or absorbing a limit cycle in the process. If we define the
bialternate matrix product of two $n\times n$ matrices $\mat{X}$ and $\mat{Y}$ to be the $\frac{1}{2}n(n-1)
\times \frac{1}{2}n(n-1)$ matrix $\mat{X} \odot \mat{Y}$ whose rows are labeled by the multi-index
$(p,q)$ (where $p=2,\ldots,n$ and $q=1,\ldots,p-1$), whose columns are labeled by the multi-index 
$(r,s)$ (where $r=2,\ldots,n$ and $s=1,\ldots,r-1$) and whose 
entries are given by

\begin{equation*}
(\mat{X} \odot \mat{Y})_{(p,q)(r,s)}
\equiv
\frac{1}{2}\left\{\left|\begin{matrix}x_{pr} & x_{ps}\\y_{qr} & y_{qs}\end{matrix}\right| + 
\left|\begin{matrix}y_{pr} & y_{ps}\\x_{qr} & x_{qs}\end{matrix}\right|\right\}
\end{equation*}

\noindent
then a necessary condition for a Hopf bifurcation to occur is \cite{guckenheimer}

\begin{equation}
\left|2\mathbf{A}\odot\mat{1}\right| = 0
\end{equation}

\noindent
with additional nondegeneracy and transversality conditions required to ensure sufficiency. 

For general CTRNNs, the pseudo-Hopf bifurcation condition $\widetilde{\mathcal{H}}$ becomes

\begin{equation}
\label{eq:H}
\widetilde{\mathcal{H}}:~\{\left|2(\mat{W}\diag(\vec{\uppsi'}) - \mat{1})\odot\mat{1}\right| 
= 0\}
\end{equation}

We call this a ``pseudo''-Hopf condition because an equilibrium point having a pair of real eigenvalues 
with equal magnitude but opposite sign (a so-called \emph{neutral saddle}) will also satisfy this condition. 
Since such points are obviously not Hopf bifurcations, the portions of $\widetilde{\mathcal{H}}$ for
which the eigenvalues are not pure imaginary must be removed in order to obtain a necessary condition
for a Hopf bifurcation $\mathcal{H}$, with additional nondegeneracy and transversality conditions required to ensure sufficiency.

\section{Local Codimension-2 Bifurcation Conditions}

\subsection{Cusp Bifurcation}

At a generic cusp (CP) bifurcation, two manifolds of codimension-1 saddle-node bifurcations coincide.
A cusp bifurcation occurs when a saddle-node bifurcation exhibits a quadratic degeneracy, so
that the local behavior is governed by the cubic term. Assuming that $\mat{A}$ has rank $n-1$, we can
define the two real vectors
$\vec{q}$ and $\vec{p}$ such that 

\begin{equation}
\label{eq:qp}
\mat{A}\vec{q}=\vec{0},~~~\mat{A}^T\vec{p}=\vec{0}
\end{equation}

\noindent
Then the quadratic coefficient of the normal form becomes

\begin{equation}
\label{eq:b}
b \equiv \frac{1}{2}\langle \vec{p}, \vec{B}(\vec{q}, \vec{q}) \rangle
\end{equation}

\noindent
and the necessary condition for a cusp bifurcation to occur is simply

\begin{equation}
\label{eq:CP}
\mathcal{CP}:~\mathcal{S} \cap \{b = 0\}
\end{equation}

\noindent 
with additional nondegeneracy and transversality conditions required for sufficiency.

For general CTRNNs, the quadratic coefficient (\ref{eq:b}) can be written as

\begin{equation}
\label{eq:CTRNN2CuspConditions}
b = \vec{p} \cdot \mat{W}\vec{\uppsi}''\vec{q}^2
\end{equation}

\subsection{Bogdanov-Takens Bifurcation}

At a generic Bogdanov-Takens (BT) bifurcation, a manifold of Hopf bifurcations branches
off from a manifold of saddle-node bifurcations. A manifold of global
homoclinic bifurcations, wherein one branch of the unstable manifold of the
saddle point coincides with one branch of its stable manifold, also arises. Such bifurcations
typically produce or absorb a large-period limit cycle.
A BT bifurcation occurs when two real eigenvalues of the Jacobian matrix $\mat{A}$
are simultaneously 0. Thus, a BT bifurcation can occur only in two or more state dimensions.

If we once again define the two real vectors $\vec{q}$ and $\vec{p}$ as in (\ref{eq:qp}), then the necessary condition for a BT bifurcation to occur is simply

\begin{equation}
\label{eq:BT}
\mathcal{BT}:~\mathcal{S} \cap \{\vec{p} \cdot \vec{q} = 0\}
\end{equation}

\noindent 
with additional nondegeneracy and transversality conditions required for sufficiency. 

A BT bifurcation can also be expressed as 

\begin{equation}
\mathcal{BT}:~\mathcal{S}\cap\mathcal{H}\cap\{\vec{p} \cdot \vec{q} = 0\}
\end{equation}

\noindent
Although this might seem redundant, it provides an interesting connection to the zero-Hopf bifurcation that we will discuss later.

\subsection{Generalized-Hopf Bifurcation}

At a generic Generalized-Hopf (GH) bifurcation (sometimes also referred to as a Bautin bifurcation), a manifold of Hopf bifurcations switches from subcritical to supercritical. A manifold of global saddle-cycle bifurcations (sometimes also called a fold bifurcation of limit cycles) also arises. In a saddle-cycle bifurcation, a pair of limit cycles of different stability coincide. A GH bifurcation occurs when the first Lyapunov coefficient of the normal form along a manifold of Hopf bifurcations passes through 0. Thus, this bifurcation can only occur in two or more state dimensions.

On a generic manifold of Hopf bifurcations $\mathcal{H}$, a single complex conjugate pair of eigenvalues is pure imaginary. If we assume that $\mat{A}$ has rank $(n-2)$, then we can define the complex vectors $\vec{q}$ and $\vec{p}$ as 

\begin{equation}
\label{eq:BTqp}
\begin{split}
\mathbf{A}\vec{q} = i \omega_0 \vec{q}&,~~~\mat{A}\overline{\vec{q}} = -i \omega_0 \overline{\vec{q}}\\
\mat{A}^T\vec{p} = -i \omega_0 \vec{p}&,~~~\mat{A}^T\overline{\vec{p}} = i \omega_0 \overline{\vec{p}}
\end{split}
\end{equation}

\noindent
where $\vec{p}$ must be normalized such that $\langle{\bf p},{\bf q}\rangle = 1$:

\begin{equation*}
    \hat{\vec{p}} \equiv \frac{\vec{p}}{\vec{p}\cdot\vec{q}}
\end{equation*}

Then the first Lyapunov coefficient $l_1$ is defined as

\begin{equation*}
\begin{split}
l_1 \equiv \frac{1}{2\omega_0}\mathrm{Re}\big[
\langle \hat{\vec{p}},\mathbf{C}(\vec{q},\vec{q},\overline{\vec{q}})\rangle - 
2\langle \hat{\vec{p}}, \mathbf{B}(\vec{q}, \mathbf{A}^{-1}\mathbf{B}(\vec{q},\overline{\vec{q}}))\rangle + 
\langle \hat{\vec{p}},  \mathbf{B}(\overline{\vec{q}}, \mat{M}^{-1}\mathbf{B}(\vec{q},\vec{q})) \rangle \big]
\end{split}
\end{equation*}

\noindent
where $\mat{M} \equiv 2i\omega_0\mat{1}-\mathbf{A}$.

The necessary condition for a GH bifurcation is then simply

\begin{equation}
\label{eq:GH}
\mathcal{GH}:~\mathcal{H} \cap \{l_1  = 0\}
\end{equation}

\noindent
with additional nondegeneracy and transversality conditions required for sufficiency.

For general CTRNNs, the first Lyapunov coefficient becomes

\begin{equation}
\label{eq:CTRNNl1}
\begin{split}
l_1 =
\frac{1}{2\omega_0}\mathrm{Re}\big[
\hat{\overline{\vec{p}}}\cdot\mat{W}\vec{q}^2\overline{\vec{q}}\boldsymbol{\uppsi}''' - 2\hat{\overline{\vec{p}}}\cdot\mat{W}\vec{q}\boldsymbol{\uppsi}''\mat{A}^{-1}\mat{W}\vec{q}\overline{\vec{q}}\boldsymbol{\uppsi}'' + \hat{\overline{\vec{p}}}\cdot\mat{W}\overline{\vec{q}}\boldsymbol{\uppsi}''\mat{M}^{-1}\mat{W}\vec{q}^2\boldsymbol{\uppsi}''
\big]
\end{split}
\end{equation}

\noindent
or, in index notation,

\begin{equation}
\label{eq:CTRNNl1_2}
\begin{split}
l_1 = \frac{1}{2\omega_0}
\bigg(
\sum_{ij}^n{w_{ij}\mathrm{Re}[\hat{\overline{p}}_i q_j^2 \overline{q}_l]\psi'''_j
+ \sum_{kl}^n{w_{ij}w_{kl}(\mathrm{Re}[\hat{\overline{p}}_i \overline{q}_j q_l^2 m^{-1}_{jk}]
- 2 a^{-1}_{jk}\mathrm{Re}[\hat{\overline{p}}_i q_j q_l \overline{q}_l])\psi''_j\psi''_l}}
\bigg)
%\sum_{ij}^n{w_{ij} \mathrm{Re}[\hat{\overline{p}}_i q_j^2 \overline{q}_l]\psi'''_j\\ 
%&+\sum_{kl}^n{w_{ij} w_{kl} (\mathrm{Re}[\hat{\overline{p}}_i \overline{q}_j q_l^2 m^{-1}_{jk}]\\
%&- 2 a^{-1}_{jk} \mathrm{Re}[\hat{\overline{p}}_i q_j q_l \overline{q}_l]) \psi''_j \psi''_l}}
%\bigg)
\end{split}
\end{equation}

\noindent
where $a^{-1}_{rs}$ and $m^{-1}_{rs}$ are the elements of $\mat{A}^{-1}$ and $\mat{M}^{-1}$, respectively,
and $\omega_0$, $\hat{\overline{p}}_r$, $q_r$, $\overline{q}_r$, $a^{-1}_{rs}$, and $m^{-1}_{rs}$ all depend on $\psi'_t$.

\subsection{Zero-Hopf Bifurcation}

At a generic Zero-Hopf (ZH) bifurcation (sometimes also called a Fold-Hopf or Gavrilov-Guckenheimer bifurcation), manifolds of saddle-node and Hopf bifurcations coincide. The complete unfolding of this codimension-2 bifurcation is complex and not yet fully understood, but it is known to contain homoclinc and heteroclinic connections, tori, and chaos. This makes a ZH bifurcation one of the few local bifurcations
that can indicate the birth of chaos. A ZH bifurcation occurs when the Jacobian matrix $\mat{A}$ simultaneously exhibits a 0 eigenvalue and a complex conjugate pair of pure imaginary eigenvalues. Thus, this bifurcation can only occur in three or more state dimensions.

The necessary condition for a ZH bifurcation to occur is

\begin{equation}
\label{eq:ZH}
    \mathcal{ZH}:~\mathcal{S}\cap\mathcal{H}\cap\{\vec{p} \cdot \vec{q} \ne 0\}
\end{equation}

\noindent
with additional nondegeneracy and transversality conditions required for sufficiency. Here the vectors
$\vec{q}$ and $\vec{p}$ are defined as in (\ref{eq:qp}). As mentioned earlier, there is
an interesting parallel between BT and ZH bifurcations. Both lie on $\mathcal{S}\cap\mathcal{H}$, with BT bifurcations occurring when $\vec{p} \cdot \vec{q} = 0$ and ZH bifurcations occurring when $\vec{p} \cdot \vec{q} \ne 0$.

\subsection{Hopf-Hopf Bifurcation}

At a generic Hopf-Hopf (HH) or double Hopf (DH) bifurcation, two independent manifolds of Hopf bifurcations
intersect. The complete unfolding of this codimension-2 bifurcation is also complex and not yet fully understood,
but, like the ZH bifurcation, is known to contain homoclinc and heteroclinic connections, tori, and chaos. An HH
bifurcation occurs when the Jacobian matrix $\mat{A}$ has two complex conjugate pairs of purely imaginary
eigenvalues and thus can only occur in four or more state dimensions.

The necessary condition for a psuedo-HH bifurcation can be written as

\begin{equation}
\widetilde{\mathcal{HH}}:~\widetilde{\mathcal{H}}\cap\widetilde{\mathcal{H}}^\perp
\end{equation}

\noindent
with additional nondegeneracy and transversality conditions required for sufficiency.
Here $\widetilde{\mathcal{H}}^\perp$ denotes the set $\{|2\mat{A}^\perp\odot\mat{1}_{n-2}|=0\}$
and $\mat{A}^\perp$ denotes the orthogonal complement of $\mat{A}$ in $\mathbb{R}^N$ w.r.t.
the 2-dimensional eigenspace associated with the first Hopf bifurcation $\mathcal{H}$. Of course,
in order to obtain $\mathcal{HH}$ from $\widetilde{\mathcal{HH}}$, both of these pesudo-Hopf bifurcations
must be further restricted so as to remove any neutral saddles.

\section{Examples}

In this section, the local codimension-1 and condimension-2 CTRNN bifurcation conditions derived above
are applied to small $\sigma$-CTRNNs containing from one to four neurons. 
In each case, we specialize the general conditions to circuits of that size, 
derive explicit expressions for those manifolds whenever possible, 
work through a concrete example circuit, 
and visually illustrate our results by presenting its parameter charts in both $\vec{\upsigma'}$ and
$\vec{\uptheta'}$ space.

\subsection{1-Neuron CTRNNs}

For a single neuron, the only local bifurcations with codimension up to two that can generically occur 
are $\mathcal{S}_1$ and $\mathcal{CP}_1$.
From (\ref{eq:SN}) and (\ref{eq:CP}), respectively, the relevant conditions can be rewritten as

\begin{equation*}
\begin{split}
\mathcal{S}_1:&~\{1 - w_{11}\sigma'_1 = 0\}\\
\mathcal{CP}^{\pm}_1:&~\{w \sigma_1'' = 0\}
\end{split}
\end{equation*}

\noindent
Using (\ref{eq:PsiPrimeSubst}), we obtain the solutions

\begin{equation*}
\begin{split}
    \mathcal{S}_1:&~\sigma_1' = \frac{1}{w_{11}},~w_{11} \ge 4\\
    \mathcal{CP}_1:&~(\sigma_1', w_{11}) = {\left(\frac{1}{4}, 4\right)}
\end{split}
\end{equation*}

\noindent
or, using (\ref{eq:SigmaTheta}) to map $\sigma_1'$ to $\theta_1$, 

\begin{equation*}
\begin{split}
 \mathcal{S}_1:&~\theta_1 = \ln{\frac{w_{11} - 2 \pm \sqrt{w_{11} (w_{11}-4)}}{2}} - \frac{w_{11} \pm \sqrt{w_{11} (w_{11} - 4)}}{2},~w_{11} \ge 4\\
 \mathcal{CP}_1:&~(\theta_1, w_{11}) = {(-2, 4)}
\end{split}
\end{equation*}

\begin{figure}[t]
\includegraphics[width=12cm]{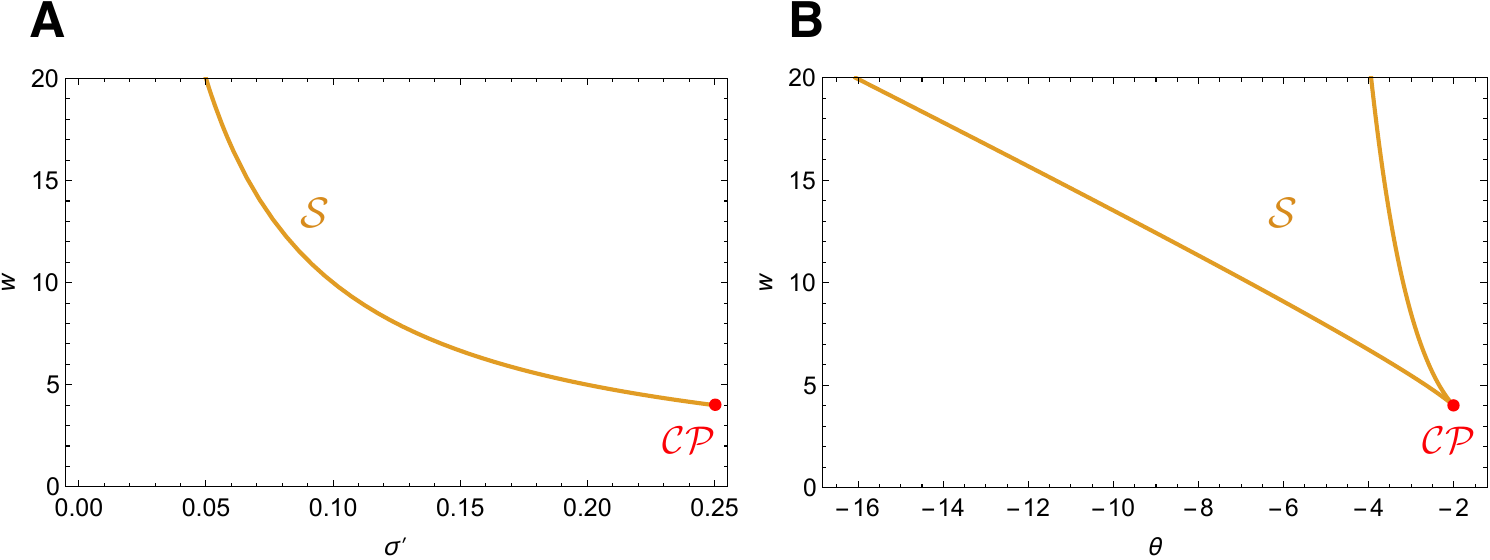}
\caption{The complete parameter chart of a 1-neuron $\sigma$-CTRNN in (A) $(\sigma',w)$-space and in (B) $(\theta,w)$-space.
Curves of saddle-node bifurcations ($\mathcal{S}$) are indicated in orange and cusp bifurcation points ($\mathcal{CP}$) are indicated by red points.}
\centering
\label{fig:CTRNN1}
\end{figure}

\noindent
These results are illustrated in Figure \ref{fig:CTRNN1}.There is no need to consider a specific
example in this case because our analysis exhaustively describes the full 2-dimensional parameter space.

\subsection{2-Neuron CTRNNs}

For a 2-neuron CTRNN, the parameter space is 6-dimensional and the bifurcations $\mathcal{S}_2$,
$\mathcal{H}_2$, $\mathcal{CP}_2$, $\mathcal{BT}_2$ and $\mathcal{GH}_2$ can occur generically.
In this section, each condition will be derived in turn and then illustrated for the example weight matrix

\begin{equation*}
\mat{W}_2 = 
\begin{pmatrix}
10 & -10\\
10 & 2
\end{pmatrix}
\end{equation*}

The saddle-node bifurcation condition in (\ref{eq:SN}) becomes 

\begin{equation*}
\mathcal{S}_2:~\{1 - w_{22}\sigma_2' - w_{11}\sigma_2' + \sigma_1' \sigma_2' \det{\mat{W}} = 0\}
\end{equation*}

\noindent
which can be solved in general (assuming nonzero denominator) to obtain

\begin{equation}
\label{eq:S2}
\mathcal{S}_2:~\sigma_2'=\frac{1-w_{11}\sigma_1'}{w_{22}-\det{\mat{W}}\sigma_1'},~0 < \sigma_1',\sigma_2' < \frac{1}{4}
\end{equation}

\begin{figure}[tp]
\includegraphics[width=12cm]{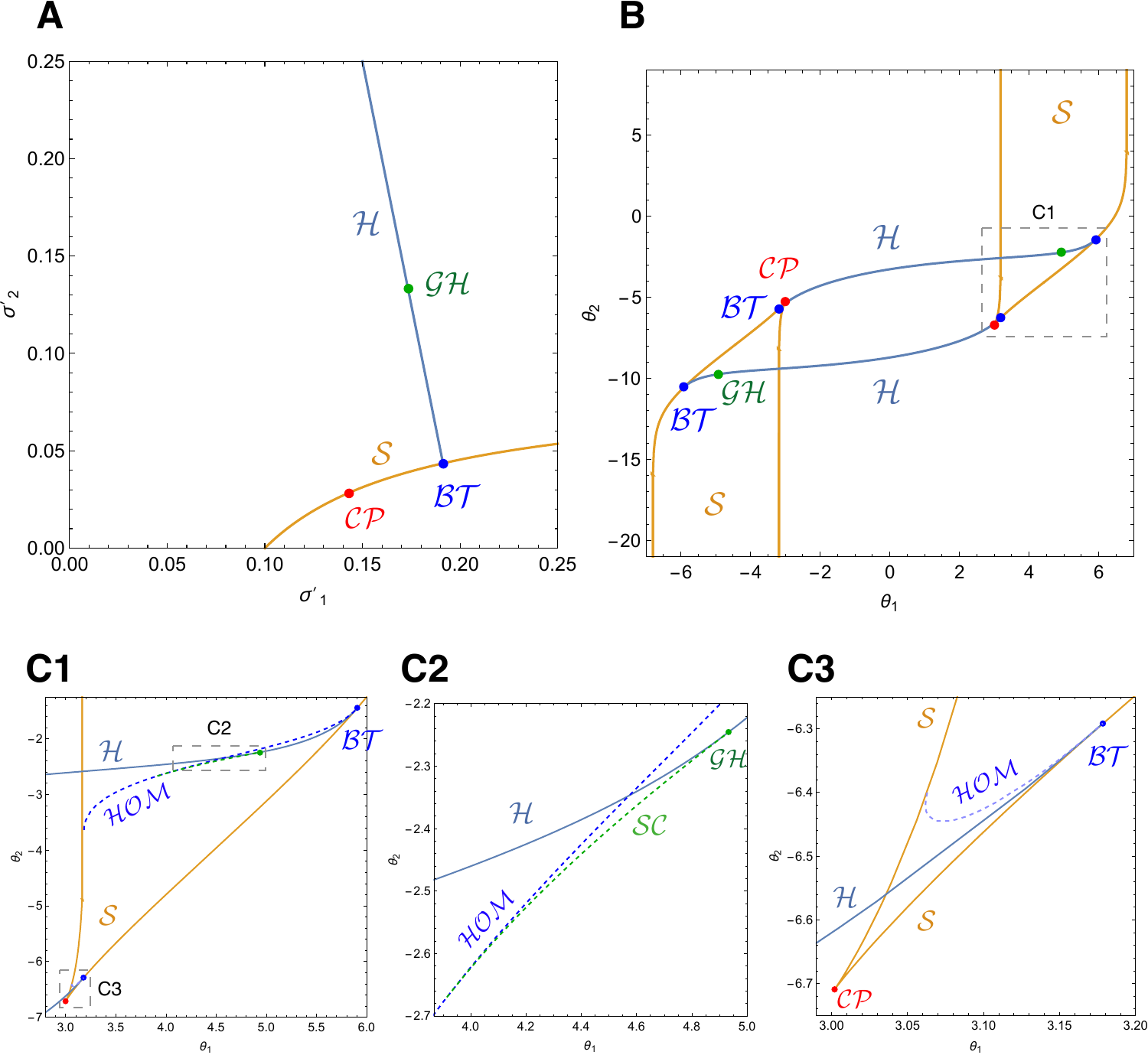}
\caption{The parameter chart of the 2-neuron $\sigma$-CTRNN given by $\mat{W}_2$ in (A) $(\sigma_1', \sigma_2')$-space and (B) $(\theta_1,\theta_2)$-space. Curves of codimension-1 saddle-node ($\mathcal{S}$) and Hopf ($\mathcal{H}$) bifurcations are shown in orange and blue, respectively. 
Codimension-2 cusp ($\mathcal{CP}$), Bogdanov-Takens ($\mathcal{BT}$) and Generalized-Hopf ($\mathcal{GH}$) bifurcations are shown as red, blue and green points, respectively. The dashed gray box in (B) is expanded in (C1) and the dashed gray boxes in (C1) are further expanded in (C2) and (C3), respectively, in order to show the curves of codimension-1 global bifurcations that arise from 
$\mathcal{BT}$ and $\mathcal{GH}$. Here the dashed lines indicate curves of homoclinic ($\mathcal{HOM}$; blue) and saddle-cycle ($\mathcal{SC}$; green) bifurcations. An analogous set of global bifurcations can be found around the other codimension-2 points in (B).}
\centering
\label{fig:CTRNN2}
\end{figure}

\noindent
For $\mat{W}_2$, this gives

\begin{equation*}
\mathcal{S}_2:~\sigma_2' = \frac{10 \sigma_1' - 1}{120 \sigma_1' - 2},~\frac{1}{10} < \sigma_1' \le \frac{1}{4}
\end{equation*}

\noindent
which is shown as an orange curve in Figure \ref{fig:CTRNN2}A.

The pseudo-Hopf bifurcation condition in (\ref{eq:H}) becomes

\begin{equation*}
\widetilde{\mathcal{H}}_2:~\{w_{11} \sigma_1' + w_{22} \sigma_2' - 2 = 0\}
\end{equation*}

\noindent
which can be solved in general (assuming $w_{22}\ne0$) and restricted to eliminate neutral saddles to obtain

\begin{equation}
\label{eq:H2}
\mathcal{H}_2:~\sigma_2' = \frac{2-w_{11}\sigma_1'}{w_{22}},~\Delta_2 < 0
\end{equation}

\noindent
For $\mat{W}_2$, this gives

\begin{equation*}
\mathcal{H}_2:~\sigma_2' = 1 - 5 \sigma_1',~\frac{6 - \sqrt{30}}{60} < \sigma_1' < \frac{6 + \sqrt{30}}{60}
\end{equation*}

\noindent
which is shown as a blue line in Figure \ref{fig:CTRNN2}A.

Next we turn to the cusp bifurcation $\mathcal{CP}_2$ in 2-neuron CTRNNs (\ref{eq:CP}).
If we assume $w_{12} w_{21} \ne 0$, the vectors $\vec{q}$ and $\vec{p}$ in (\ref{eq:qp}) can be written as

\begin{equation*}
\vec{q} = (w_{22} \sigma_2' - 1, -w_{21} \sigma_1'),~~~\vec{p} = (w_{22} \sigma_2' - 1, -w_{12} \sigma_2')
\end{equation*}

\noindent
and the quadratic coefficient (\ref{eq:CTRNN2CuspConditions}) becomes

\begin{equation*}
b=(w_{22}\sigma_2'-1)^2 (\det{\mat{W}}\sigma_2'-w_{11})\sigma_1'' - w_{12}w_{21}^2 \sigma_1'^2 \sigma_2''
\end{equation*}

\noindent
Using (\ref{eq:PsiPrimeSubst}), $\sigma''$ can be rewritten in terms of $\sigma'$, 
with the different possible sign combinations splitting $b$ into four expressions,
only two of which are distinct:

\begin{equation*}
    \begin{split}
         b^{\pm\pm} =& (w_{22} \sigma_1' - 1)^2 (\det{\mat{W}} \sigma_2' - w_{11}) \sigma_1' \sqrt{1 - 4 \sigma_1'} - w_{12} w_{21}^2 {\sigma_1'}^2 \sigma_2' \sqrt{1 - 4 \sigma_2'}\\
         b^{\pm\mp} =& (w_{22} \sigma_2' - 1)^2 (\det{\mat{W}} \sigma_2' - w_{11}) \sigma_1' \sqrt{1 - 4 \sigma_1'} + w_{12} w_{21}^2 {\sigma_1'}^2 \sigma_2' \sqrt{1 - 4 \sigma_2'}
    \end{split}
\end{equation*}

\noindent
This in turn give us the cusp manifolds

\begin{equation}
\label{eq:CP2}
\begin{split}
\mathcal{CP}_2^{\pm\pm}:&~\mathcal{S}_2 \cap \{b^{\pm\pm} = 0\}\\
\mathcal{CP}_2^{\pm\mp}:&~\mathcal{S}_2 \cap \{b^{\pm\mp} = 0\}
\end{split}
\end{equation}

\noindent
In general, these equations can only be solved in terms of the roots of a certain 6th-order polynomial. 
Thus, we must typically resort to numerical solutions. For $\mat{W}_2$, this gives

\begin{equation*}
\mathcal{CP}_2^{\pm\pm}:~(\sigma_1', \sigma_2') = (0.143151, 0.0284297)
\end{equation*}

\noindent
which appears as a red point in Figure \ref{fig:CTRNN2}A. Note that, for $\mat{W}_2$, the $\mathcal{CP}_2^{\pm\mp}$ 
condition does not have any real solutions within the constraints on 
$\sigma_1'$ and $\sigma_2'$.

The Bogdanov-Takens bifurcation condition $\mathcal{BT}_2$ in (\ref{eq:BT}) becomes

\begin{equation*}
\mathcal{BT}_2:~\mathcal{S}_2 \cap \{w_{12}^2 \sigma_2'^2 + (w_{22}\sigma_2' - 1)^2 = 0\}
\end{equation*}

\noindent
These equations can be solved explicitly to obtain

\begin{equation}
\label{eq:BT2}
\mathcal{BT}_2:~(\sigma_1', \sigma_2') = 
\Biggl\{\left(\frac{w_{12}w_{21}+\alpha}{\alpha w_{11}}, \frac{w_{11}}{\det{\mat{W}}-\alpha}\right),
\left(\frac{w_{22}}{\det{\mat{W}}-\alpha}, \frac{w_{12}w_{21}+\alpha}{\alpha w_{22}}\right)\Bigg\}
\end{equation}

\noindent
where the abbreviation $\alpha \equiv \sqrt{-w_{12}w_{21}\det{\mat{W}}}$.
Note that these solutions only exist when
$\alpha$ is real, $w_{11}, w_{22} \ne 0$ and $0 < \sigma_1', \sigma_2' \le 1/4$.
For $\mat{W}_2$, this gives

\begin{equation*}
\mathcal{BT}_2:~ (\sigma_1', \sigma_2') = \left(\frac{6 + \sqrt{30}}{60}, \frac{6 - \sqrt{30}}{60}\right)
\end{equation*}

\noindent
which is shown as a blue point in Figure \ref{fig:CTRNN2}A.

Finally, we consider the Generalized Hopf bifurcation condition $\mathcal{GH}_2$ in (\ref{eq:GH}).
For generic 2-neuron CTRNNs satisfying the Hopf condition $\mathcal{H}_2$ in (\ref{eq:H2}), the eigenvalues take
the form

\begin{equation*}
\lambda_1, \lambda_2 = \pm i \omega_0
\end{equation*}

\noindent
where

\begin{equation*}
\omega_0 \equiv \sqrt{-(1 - 2 w_{22} \sigma_2' + w_{12} w_{21} \sigma_1' \sigma_2' + w_{22}^2 \sigma_2'^2)} 
\end{equation*}

\noindent
Assuming that $w_{21} \ne 0$ the corresponding eigenvectors are

\begin{equation*}
\vec{v}_1, \vec{v}_2 = (1 - w_{22} \sigma_2' \pm i \omega_0, w_{21} \sigma_1')
\end{equation*}

\noindent
A similar analysis can be applied to $\mat{A}^T$, with $w_{12} \ne 0$. Comparing these results to (\ref{eq:BTqp}), we find that the vectors $\vec{q}$ and $\vec{p}$ can be written as

\begin{equation*}
\begin{split}
\vec{q} &= (1 - w_{22} \sigma_2' + i \omega_0, w_{21} \sigma_1')\\
\vec{p} &= (1 - w_{22} \sigma_2' - i \omega_0, w_{12} \sigma_2')
\end{split}
\end{equation*}

\noindent
We can substitute these vectors into (\ref{eq:CTRNNl1}), symbolically extract the real part, use (\ref{eq:PsiPrimeSubst}) and (\ref{eq:PsiPrimePrimeSubst}) to rewrite the higher
derivatives $\sigma''$ and $\sigma'''$ that appear, eliminate $\sigma_2'$ by restricting to the manifold $\mathcal{H}_2$, and then algebraically simplifying the result to obtain 

\begin{align}
\label{eq:CTRNNGH2}
l_1 = &(w_{21} \sigma_1'^2 (w_{11} \sigma_1'-2) (w_{21} (w_{22} (-2 \alpha s w_{12}+w_{22}-4)+8 w_{12} w_{21} \sigma_1') 
\notag\\&~+w_{11} (w_{12} (w_{22} (2 \sigma_1' (\alpha s w_{21}-3)+1)-8 w_{21}^2 \sigma_1'^2)-4 w_{12}{}^2 w_{21} \sigma_1'^2 
\notag\\&~-w_{21} (w_{22}-2) w_{22} \sigma_1')-2 w_{11}^3 (w_{12}+w_{21}) w_{22} \sigma_1'^3 
\notag\\&~+w_{11}^2 \sigma_1' (2 w_{12} w_{21} (w_{12}+w_{21}) \sigma_1'^2+4 (2 w_{12}+w_{21}) w_{22} \sigma_1'-w_{12} w_{22})))
\notag\\& /(4 \beta  w_{22} (w_{22} (w_{11} \sigma_1'-1)^2+w_{12} w_{21} \sigma_1' (2-w_{11} \sigma_1')))
\end{align}
  
\noindent
where we have utilized the abbreviations

\begin{align*}
\alpha \equiv &\sqrt{1-4 \sigma_1'} \sqrt{\frac{4 w_{11} \sigma_1'+w_{22}-8}{w_{22}}}\\
\beta \equiv &\sqrt{\frac{(w_{12} w_{21}-w_{11} w_{22}) \sigma_1' (w_{11} \sigma_1'-2)}{w_{22}}-1}
\end{align*}

\noindent
and $s = \pm1$ depending on whether we choose the same or opposite branches for $\sigma_1''$ and $\sigma_2''$ in
(\ref{eq:PsiPrimeSubst}). This leads to two expressions for the first Lyapunov coefficient, $l_1^{\pm\pm}$ and $l_1^{\pm\mp}$, and hence two systems of equations determining generalized Hopf bifurcations in 2-neuron $\sigma$-CTRNNs 

\begin{equation}
\label{eq:GH2}
\begin{split}
\mathcal{GH}^{\pm\pm}_2:&~\mathcal{H}_2 \cap \{l_1^{\pm\pm} = 0 \}\\
\mathcal{GH}^{\pm\mp}_2:&~\mathcal{H}_2 \cap \{l_1^{\pm\mp} = 0 \}
\end{split}
\end{equation}

\noindent
The $l_1$ equations can be solved for $\sigma_1'$ (numerically if necessary) and then $\sigma_2'$
can be computed from $\mathcal{H}_2$ as $\sigma_2' = (2-w_{11}\sigma_1')/w_{22}$.  
As usual, only real solutions for which $0 < \sigma_1', \sigma_2' \le 1/4$ should be considered.
For $\mat{W}_2$, this gives

\begin{equation*}
\mathcal{GH}^{\pm\mp}_2:~ (\sigma_1', \sigma_2') = 
\left(\frac{\sqrt{85} - \sqrt{4\sqrt{85} - 31} + 14}{120}, \frac{\sqrt{4\sqrt{85} - 31} - 10}{24}\right)
\end{equation*}

\noindent
which is shown as a green point in Figure \ref{fig:CTRNN2}A.

Finally, all of the bifurcation manifolds derived above can be mapped from activation function derivative space
$(\sigma_1', \sigma_2')$ to net input space $(\theta_1, \theta_2)$ using (\ref{eq:SigmaTheta}) and
(\ref{eq:DSigmaInverse}). In doing so, note that care must be taken to maintain branch consistency. For example, while
$\mathcal{BT}_2$ in $\vec{\upsigma}'$-space gives rise to the four points $\vec{\Uptheta}^{++}(\mathcal{BT}_2)$,
$\vec{\Uptheta}^{+-}(\mathcal{BT}_2)$, $\vec{\Uptheta}^{-+}(\mathcal{BT}_2)$, and
$\vec{\Uptheta}^{--}(\mathcal{BT}_2)$ in $\vec{\uptheta}$-space, $\mathcal{GH}^{\pm\mp}_2$ only gives rise to 
the two points $\vec{\Uptheta}^{+-}(\mathcal{GH}^{+-}_2)$ and $\vec{\Uptheta}^{-+}(\mathcal{GH}^{-+}_2)$.
These results are illustrated in Figure \ref{fig:CTRNN2}B.

Having calculated the locations of all the codimension-2 bifurcation points in $\mat{W}_2$, we can now use a
normal form theory analysis to approximate the direction and initial shape of the homoclinic and saddle-cycle global bifurcation curves that emanate from the BT and GH bifurcation points, respectively
\cite{kuznetsov}. These codimension-1 bifurcations can then be followed using numerical continuation, resulting
in the dashed curves shown in Figure \ref{fig:CTRNN2}C.

In the example above, bifurcation manifolds were derived for fixed weights. However, nothing prevents us from
applying the same analysis to 2-neuron CTRNNs containing one or more free weights.
For example, consider the following generalization of $\mat{W}_2$ with one free weight $w$

\begin{equation*}
\mat{W}'_2 = 
\begin{pmatrix}
10 & w\\
10 & 2
\end{pmatrix}
\end{equation*}

\noindent
Applying the conditions (\ref{eq:S2}-\ref{eq:GH2}) to this weight matrix gives

\begin{equation*}
\begin{split}
\mathcal{S}_2:&~ w = \frac{1 - 10 \sigma_1' - 2 \sigma_2' + 20 \sigma_1' \sigma_2'}{10 \sigma_1' \sigma_2'}\\
\mathcal{H}_2:&~ \sigma_2' = 1 - 5 \sigma_1',~\frac{3}{20} \le \sigma_1' < \frac{1}{10} + \frac{1}{10}\sqrt{\frac{w}{w-2}},~ w < -\frac{2}{3}\\ 
%
%\mathcal{CP}^{\pm\pm}_2:&~ \Bigg\{(\sigma_1', \sigma_2', w) \Bigg\lvert \frac{1}{2}\sqrt{\frac{5(w+2)\sigma_1'-1}{5(w-2)\sigma_1'+1}}-\frac{5w^2\sigma_1'\sqrt{1-4\sigma_1'}}{(40\sigma_1'-4)(1+5(w-2)\sigma_1')^2} = 0,~\sigma_2' = \frac{1-10\sigma_1'}{2-20\sigma_1'+10w\sigma_1'}\Bigg\}\\ 
\mathcal{CP}^{\pm\pm}_2:&~ \Bigg\{ \frac{1}{2}\sqrt{\frac{5(w+2)\sigma_1'-1}{5(w-2)\sigma_1'+1}}-\frac{5w^2\sigma_1'\sqrt{1-4\sigma_1'}}{(40\sigma_1'-4)(1+5(w-2)\sigma_1')^2} = 0,~\sigma_2' = \frac{1-10\sigma_1'}{2-20\sigma_1'+10w\sigma_1'}\Bigg\}\\ 
\mathcal{BT}_2:&~ (\sigma_1', \sigma_2') = \left(-\frac{1}{5(w-2+\sqrt{w(w-2)})}, \frac{1}{2-w+\sqrt{w(w-2)}},\right),~w \le -\frac{2}{3}\\ 
\mathcal{GH}^{\pm\pm}_2:&~(\sigma_2', w) = \left(1 - 5 \sigma_1', \frac{\alpha+\gamma+\sqrt{\beta+\delta^+}}{40\sigma_1'^2(5\sigma_1'-1)}\right),~\frac{3}{20} \le \sigma_1' < \frac{1}{10} + \frac{1}{10}\sqrt{\frac{w}{w-2}},~ w < -\frac{2}{3}\\
\mathcal{GH}^{\pm\mp}_2:&~(\sigma_2', w) = \left(1 - 5 \sigma_1', \frac{\alpha-\gamma+\sqrt{\beta+\delta^-}}{40\sigma_1'^2(5\sigma_1'-1)}\right),~\frac{3}{20} \le \sigma_1' < \frac{1}{10} + \frac{1}{10}\sqrt{\frac{w}{w-2}},~ w < -\frac{2}{3}
\end{split}
\end{equation*}

\noindent
where we have used the following abbreviations

\begin{equation*}
\begin{split}
\alpha \equiv&~ -800\sigma_1'^3 + 320 \sigma_1'^2 - 24\sigma_1' - 1 \\
\beta \equiv&~ 800000\sigma_1'^6-320000\sigma_1'^5+32000\sigma_1'^4+800\sigma_1'^3-160\sigma_1'^2\\
\gamma \equiv&~ (2-20\sigma_1')\sqrt{1-4\sigma_1'}\sqrt{20\sigma_1'-3}\\
\delta^+ \equiv&~ \left(1 - 2\sqrt{1-4\sigma_1'}\sqrt{20\sigma_1'-3} + 4\sigma_1'\left(6 + 40\sigma_1'(5\sigma_1'-2) + 5\sqrt{1-4\sigma_1'}\sqrt{20\sigma_1'-3}\right)\right)^2\\
\delta^- \equiv&~ \left(1 + 2\sqrt{1-4\sigma_1'}\sqrt{20\sigma_1'-3} + 4\sigma_1'\left(6 + 40\sigma_1'(5\sigma_1'-2) - 5\sqrt{1-4\sigma_1'}\sqrt{20\sigma_1'-3}\right)\right)^2
\end{split}
\end{equation*}

\begin{figure}[t]
\includegraphics[width=12cm]{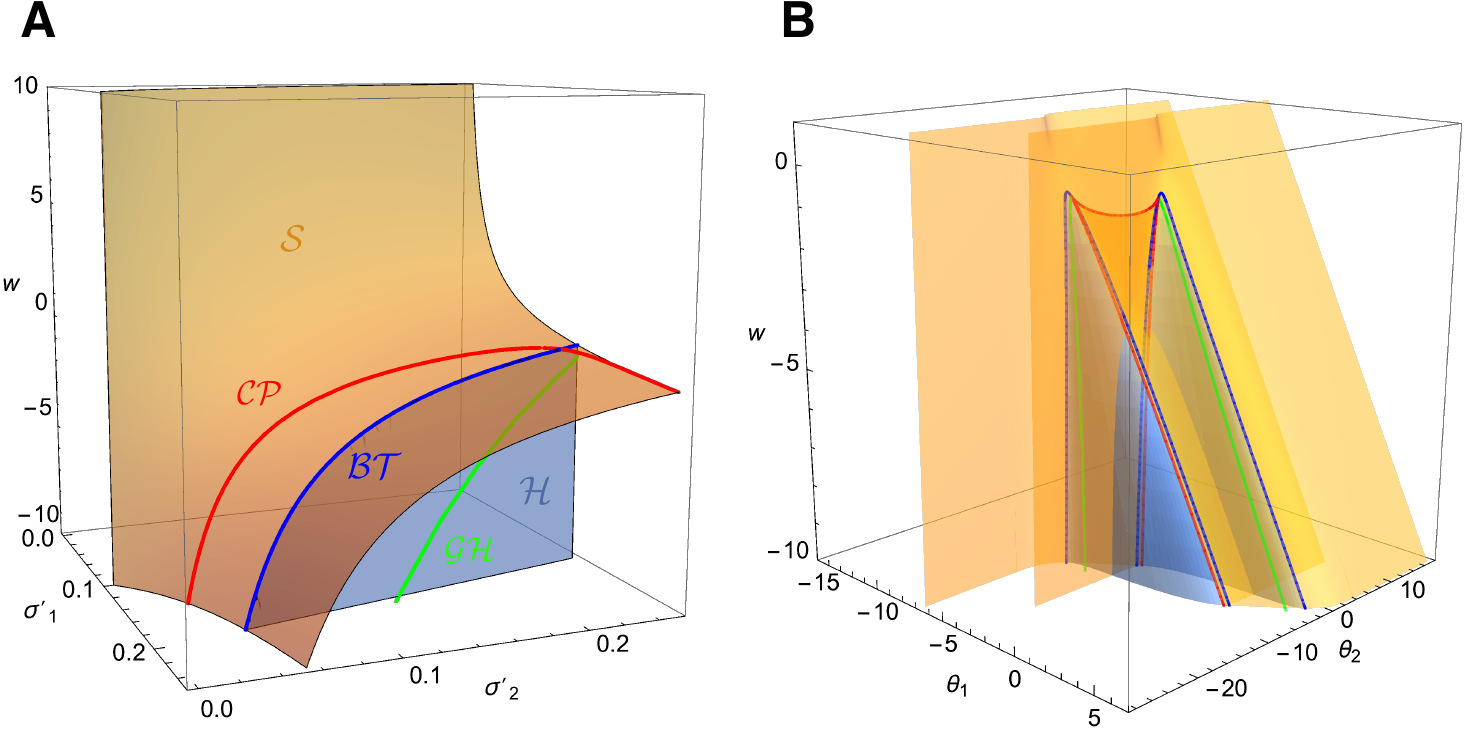}
\caption{The parameter chart of the 2-neuron $\sigma$-CTRNN with one free
weight given by $\mat{W}'_2$ in (A) ($\sigma_1', \sigma_2', w$)-space and (B) ($\theta_1,\theta_2, w$)-space. Surfaces of codimension-1 saddle-node ($\mathcal{S}$) and Hopf ($\mathcal{H}$) bifurcations are shown in orange and blue, respectively. 
Codimension-2 cusp ($\mathcal{CP}$), Bogdanov-Takens ($\mathcal{BT}$) and Generalized-Hopf ($\mathcal{GH}$) bifurcations are shown as red, blue and green curves, respectively.}
\centering
\label{fig:CTRNN2w}
\end{figure}

\noindent
It turns out that $\mathcal{CP}^{\pm\pm}_2$ can actually also be solved for $w$ in closed form, but the solution
is too complicated to reproduce here. These bifurcation manifolds are shown in Figure \ref{fig:CTRNN2w}A.
Finally, all of the bifurcation manifolds derived above can be mapped from activation function 
derivative space $(\sigma_1',
\sigma_2')$ to net input space $(\theta_1, \theta_2)$ using (\ref{eq:SigmaTheta}) and
(\ref{eq:DSigmaInverse}), taking care to maintain branch consistency (Figure \ref{fig:CTRNN2w}B).
Note that the codimension-2 curves in this circuit come together at higher-codimension points near the
top of the figure.

\subsection{3-Neuron CTRNNs}

For a 3-neuron CTRNN, the parameter space is 12-dimensional and the bifurcations $\mathcal{S}_3$,
$\mathcal{H}_3$, $\mathcal{CP}_3$, $\mathcal{BT}_3$, $\mathcal{GH}_3$ and $\mathcal{ZH}_3$ 
can occur generically. In this section, each condition will be derived in turn and then illustrated 
for the example weight matrix

\begin{equation*}
\mat{W}_3 = 
\begin{pmatrix}
6 & -1 & 1\\
1 & 6  & -1\\
-1& 1  & 6
\end{pmatrix}
\end{equation*}

The saddle-node bifurcation condition in (\ref{eq:SN}) becomes

\begin{equation*}
\mathcal{S}_3:~\{-1+w_{11}\sigma_1'+w_{22}\sigma_2'+w_{33}\sigma_3' - \det{\mat{W}_{33}}\sigma_1'\sigma_2' - \det{\mat{W}_{22}}\sigma_1'\sigma_3' - \det{\mat{W}_{11}}\sigma_2'\sigma_3' + \det{\mat{W}}\sigma_1'\sigma_2'\sigma_3' = 0\}
\end{equation*}

\noindent
which can be solved in general (assuming nonzero denominator) to obtain

\begin{equation}
\mathcal{S}_3:~\sigma_3'=\frac{1-w_{11}\sigma_1'-w_{22}\sigma_2'+\det{\mat{W}_{33}}\sigma_1'\sigma_2'}{w_{33}-\det{\mat{W}_{22}}\sigma_1'-\det{\mat{W}_{11}}\sigma_2'+\det{\mat{W}}\sigma_1'\sigma_2'},~0 < \sigma_1',\sigma_2',\sigma_3' < \frac{1}{4}
\end{equation}

\begin{figure}[t!]
\includegraphics[width=12cm]{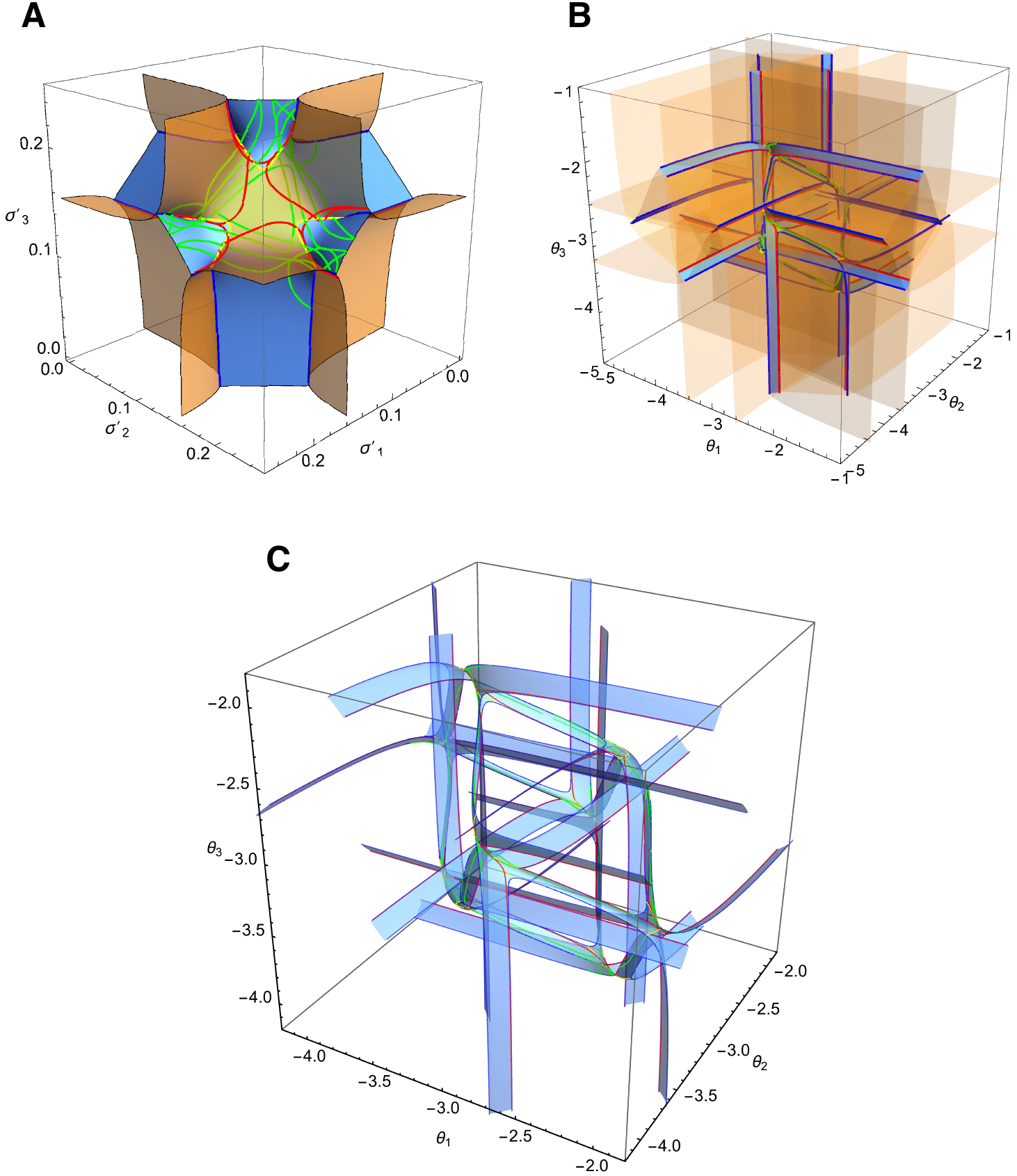}
\caption{The parameter chart of the 3-neuron $\sigma$-CTRNN given by $\mat{W}_3$ in (A) $\vec{\upsigma}'$-space and (B) $\vec{\uptheta}$-space. Surfaces of codimension-1 saddle-node ($\mathcal{S}$) and Hopf ($\mathcal{H}$) bifurcations are shown in orange and blue, respectively. 
Codimension-2 cusp ($\mathcal{CP}$), Bogdanov-Takens $(\mathcal{BT})$, Generalized-Hopf ($\mathcal{GH}$), and Zero-Hopf ($\mathcal{ZH}$) bifurcations are shown as red, blue, green and yellow curves, respectively. (C) A close-up of the central portion of (B) with the saddle-node bifurcation surfaces removed for clarity.}
\centering
\label{fig:CTRNN3}
\end{figure}

\noindent
For $\mat{W}_3$, this gives

\begin{equation*}
\mathcal{S}_3:~\sigma_3'=\frac{1-6\sigma_1'-6\sigma_2'+37\sigma_1'\sigma_2'}{6-37\sigma_1'-37\sigma_2'+234\sigma_1'\sigma_2'},~0 < \sigma_1',\sigma_2',\sigma_3' < \frac{1}{4}
\end{equation*}

\noindent
which is shown as an orange surface in Figure \ref{fig:CTRNN3}A. Note that the restrictions lead to a
somewhat complicated region of definition for these expressions.

The pseudo-Hopf bifurcation condition in (\ref{eq:H}) becomes

\begin{equation*}
\begin{split}
\widetilde{\mathcal{H}}_3:~&\{-w_{21}\sigma_1'\sigma_2'(w_{12}(w_{11}\sigma_1'+w_{22}\sigma_2'-2)+w_{13}w_{23}\sigma_3')\\
&-w_{31}\sigma_1'\sigma_3'(w_{13}(w_{11}\sigma_1'+w_{33}\sigma_3'-2)+w_{12}w_{23}\sigma_2')\\
&+(w_{22}\sigma_2'+w_{33}\sigma_3'-2) (-w_{23}w_{32}\sigma_2'\sigma_3'+(w_{11}\sigma_1'+w_{22}\sigma_2'-2) (w_{11}\sigma_1'+w_{33}\sigma_3'-2)) = 0\}
\end{split}
\end{equation*}

\noindent
which can be solved in general (assuming nonzero denominator) and restricted to eliminate neutral
saddles to obtain

\begin{equation}
\label{eq:H3}
\mathcal{H}_3:~\sigma_3' = \left\{\frac{\alpha+\sqrt{\beta}}{\gamma},\frac{\alpha-\sqrt{\beta}}{\gamma}\right\},~\Delta_3<0,~0 < \sigma_1',\sigma_2',\sigma_3' < \frac{1}{4}
\end{equation}

\noindent
where

\begin{equation*}
\begin{split}
\alpha \equiv&~(6w_{11}w_{33}-2w_{13}w_{31})\sigma_1' + (w_{11}w_{13}w_{31}-w_{11}^2w_{33})\sigma_1'^2\\&~+ (6w_{22} w_{33}-2w_{23}w_{32})\sigma_2' + 
 (w_{12}w_{23}w_{31}+w_{13}w_{21}w_{32}-2w_{11}w_{22}w_{33})\sigma_1'\sigma_2'\\&~+ (w_{22}w_{23}w_{32}-w_{22}^2w_{33})\sigma_2'^2 - 8w_{33}\\
\beta \equiv&~-4w_{33}(w_{11}\sigma_1'+w_{22}\sigma_2'-2)(4-2w_{22} \sigma_2'-
    w_{12}w_{21}\sigma_1' \sigma_2'+w_{11}\sigma_1'(w_{22}\sigma_2'-2))\\&~\times (-w_{13}w_{31}\sigma_1'- 
    w_{23}w_{32} \sigma_2'+ 
    w_{33}(w_{11}\sigma_1'+w_{22}\sigma_2'-2))\\&~ + (-w_{33}( 
      w_{11}\sigma_1'+w_{22}\sigma_2'-4)(w_{11}\sigma_1'+w_{22}\sigma_2'-2)\\&~~~~~+ 
   w_{13}\sigma_1'(w_{11}w_{31}\sigma_1'+w_{21}w_{32}\sigma_2'-2w_{31})\\&~~~~~+ 
   w_{23}\sigma_2'(w_{12}w_{31}\sigma_1'+w_{22}w_{32}\sigma_2'-2w_{32}))^2\\
\gamma \equiv&~(2 w_{11} w_{33}^2 - 2 w_{13} w_{31} w_{33}) \sigma_1' + (2 w_{22} w_{33}^2 - 2 w_{23} w_{32} w_{33}) \sigma_2' - 4 w_{33}^2
\end{split}
\end{equation*}

\clearpage
\noindent
For $\mat{W}_3$, these abbreviations become

\begin{equation*}
\begin{split}
\alpha \equiv&~218\sigma_1' - 222\sigma_1'^2 + 218\sigma_2' - 
 432\sigma_1' \sigma_2' - 222\sigma_2'^2 - 48\\
\beta \equiv&~49284\sigma_1'^4-32856\sigma_1'^3+5476\sigma_1'^2-5328\sigma_1'^3\sigma_2'+36312\sigma_1'^2\sigma_2'-109080\sigma_1'^2\sigma_2'^2\\&~-11512\sigma_1'\sigma_2'+36312\sigma_1'\sigma_2'^2-5328\sigma_1'\sigma_2'^3+5476\sigma_2'^2-32856\sigma_2'^3+49284\sigma_2'^4\\
\gamma \equiv&~ 444 \sigma_1' + 444 \sigma_2' - 144
\end{split}
\end{equation*}

\noindent
giving rise to the blue surfaces in Figure \ref{fig:CTRNN3}A. Note that the restrictions lead to a complicated
region of definition for these expressions.

Next we turn to the cusp bifurcation $\mathcal{CP}_3$ in 3-neuron CTRNNs (\ref{eq:CP}).
If we assume $\det{\mat{W}_{13}}\sigma'_1\sigma'_2 + w_{31}\sigma'_1 \ne 0$ and $\det{\mat{W}_{31}}\sigma'_2\sigma'_3 + w_{13}\sigma'_3 \ne 0$, the vectors $\vec{q}$ and $\vec{p}$ in (\ref{eq:qp}) can be written as

\begin{equation}
\begin{split}
\label{eq:qp3}
{\small \vec{q} = (-\det{\mat{W}_{11}}\sigma'_2\sigma'_3+w_{22}\sigma'_2+w_{33}\sigma'_3-1, \det{\mat{W}_{12}}\sigma'_1\sigma'_3-w_{21}\sigma'_1, -\det{\mat{W}_{13}}\sigma'_1\sigma'_2-w_{31}\sigma'_1)}\\
{\small \vec{p} = (-\det{\mat{W}_{11}}\sigma'_2\sigma'_3+w_{22}\sigma'_2+w_{33}\sigma'_3-1, \det{\mat{W}_{21}}\sigma'_2\sigma'_3-w_{12}\sigma'_2, -\det{\mat{W}_{31}}\sigma'_2\sigma'_3 - w_{13}\sigma'_3)}
\end{split}
\end{equation}

\noindent
and the quadratic coefficient (\ref{eq:b}) becomes

\begin{equation*}
\begin{split}
b=&~(1-w_{33}\sigma_3'-\sigma_2'(w_{22}-\det{\mat{W}_{11}}\sigma_3'))^2(w_{11}-\det{\mat{W}_{22}}\sigma_3'-\sigma_2'(\det{\mat{W}_{33}}-\det{\mat{W}}\sigma_3'))\sigma_1''\\&~-\sigma_1'^2(w_{21}-\det{\mat{W}_{12}}\sigma_3')^2(w_{12}-\det{\mat{W}_{21}}\sigma_3')\sigma_2''\\&~-\sigma_1'^2(w_{31}-\det{\mat{W}_{13}}\sigma_2')^2(w_{13}-\det{\mat{W}_{31}}\sigma_2')\sigma_3''\\
\end{split}
\end{equation*}

\noindent
The $\sigma_i''$ terms in this expression can be rewritten in terms of $\sigma_i'$ using (\ref{eq:PsiPrimeSubst}).
Since each $\sigma_i''$ can have either sign, there are a total of eight ways to accomplish this, only four of which are distinct, giving rise to up to four cusp bifurcation manifolds:

\begin{equation}
\begin{split}
\mathcal{CP}_3^{\pm\pm\pm}:&~\mathcal{S}_3 \cap \{b^{\pm\pm\pm} = 0\}\\
\mathcal{CP}_3^{\pm\pm\mp}:&~\mathcal{S}_3 \cap \{b^{\pm\pm\mp} = 0\}\\
\mathcal{CP}_3^{\pm\mp\pm}:&~\mathcal{S}_3 \cap \{b^{\pm\mp\pm} = 0\}\\
\mathcal{CP}_3^{\pm\mp\mp}:&~\mathcal{S}_3 \cap \{b^{\pm\mp\mp} = 0\}
\end{split}
\end{equation}

\noindent
In general, the resulting systems of equations must be solved numerically.
For $\mat{W}_3$, the quadratic coefficient $b$ becomes

\begin{equation*}
\begin{split}
b=&~(1-6\sigma_3'-\sigma_2'(6-37\sigma_3'))^2(6-37\sigma_3'-\sigma_2'(37-234\sigma_3'))\sigma_1''\\&~-\sigma_1'^2(1-5\sigma_3')^2(-1+7\sigma_3')\sigma_2''\\&~-\sigma_1'^2(-1-7\sigma_2')^2(1+5\sigma_2')\sigma_3''
\end{split}
\end{equation*}

\noindent
producing the CP bifurcation curves shown in red in Figure \ref{fig:CTRNN3}A. Note that there is a singularity
that must be avoided at $(\sigma_1',\sigma_2')=(1/7,1/5)$. This arises from violations of the conditions on the
derivations of $\vec{q}$ and $\vec{p}$ in (\ref{eq:qp3}).

For BT bifurcations, we can use the same expressions for the vectors $\vec{q}$ and $\vec{p}$ as given in
(\ref{eq:qp3}), with the same stated assumptions applying. In this case, the BT condition becomes

\begin{equation}
\begin{split}
\mathcal{BT}_3:~\mathcal{S}_3 \cap
\{&\sigma'_1\sigma'_3(w_{13} +\det{\mat{W}_{31}}\sigma'_2)(w_{31} + \det{\mat{W}_{13}}\sigma'_2) \\&+ \sigma'_1\sigma'_2(w_{12} - \det{\mat{W}_{21}}\sigma'_3)(w_{21} - \det{\mat{W}_{12}}\sigma) \\
&+ (1 - w_{33}\sigma'_3 + \sigma'_2 (-w_{22} + \det{\mat{W}_{11}}\sigma'_3))^2 = 0\}
\end{split}
\end{equation}

\noindent
This set of equations can actually be solved explicitly by
first solving $\mathcal{S}_3$ for $\sigma_3'$ and then substituting that solution into 
$\vec{p}\cdot\vec{q} = 0$ and solving for $\sigma_2'$, giving a bifurcation manifold parameterized by
$\sigma_1'$. However, the resulting general solution is quite complicated and we do not reproduce it here.
For $\mat{W}_3$, this gives the solutions

\begin{equation*}
\begin{split}
\mathcal{BT}_3:~(\sigma_1',\sigma_2',\sigma_3') =& \left\{\left(\sigma_1', \frac{\alpha+\sqrt{\beta}}{\gamma}, \frac{\alpha-\sqrt{\beta}}{\gamma}\right), \left(\sigma_1', \frac{\alpha-\sqrt{\beta}}{\gamma}, \frac{\alpha+\sqrt{\beta}}{\gamma}\right)\right\},\\
&0<\sigma_1'<\frac{74-\sqrt{97}}{489} ~~\textrm{or}~~\\
&\frac{37-\sqrt{37}}{122}<\sigma_1'<\frac{74+\sqrt{97}}{489} ~~\textrm{or}~~\\
&\frac{1+\sqrt{37}}{6\sqrt{37}}<\sigma_1'<\frac{1}{4} 
\end{split}
\end{equation*}

\noindent
where

\begin{equation*}
\begin{split}
\alpha \equiv&~37+6\sigma_1'(234\sigma_1'-77)\\
\beta \equiv&~37+12\sigma_1'(-76+3\sigma_1'(231+13\sigma_1'(105\sigma_1'-71)))\\
\gamma \equiv&~6(37+39\sigma_1'(37\sigma_1'-12))
\end{split}
\end{equation*}

\noindent
These solutions give rise to the blue curves of BT bifurcations shown in Figure ~\ref{fig:CTRNN3}A.

For GH bifurcations, it is possible in principle to repeat the derivations carried out above for
the 2-neuron case. In practice, however, it is difficult to symbolically determine which pair of 
eigenvalues are purely complex on $\mathcal{H}_3$ and, even if we could, the resulting expressions for
$l_1$ become extremely complicated. Thus, we must resort to numerically determining $\vec{q}$ and $\vec{p}$ and then expanding and numerically solving (\ref{eq:CTRNNl1}) or (\ref{eq:CTRNNl1_2}) on
$\mathcal{H}_3$ to obtain

\begin{equation}
\begin{split}
\mathcal{GH}_3^{\pm\pm\pm}:&~\mathcal{H}_3 \cap \{l_1^{\pm\pm\pm} = 0\}\\
\mathcal{GH}_3^{\pm\pm\mp}:&~\mathcal{H}_3 \cap \{l_1^{\pm\pm\mp} = 0\}\\
\mathcal{GH}_3^{\pm\mp\pm}:&~\mathcal{H}_3 \cap \{l_1^{\pm\mp\pm} = 0\}\\
\mathcal{GH}_3^{\pm\mp\mp}:&~\mathcal{H}_3 \cap \{l_1^{\pm\mp\mp} = 0\}
\end{split}
\end{equation}

\noindent
For $\mat{W}_3$, this gives rise to the complicated network of green GH bifurcation curves shown in Figure \ref{fig:CTRNN3}A.

Finally, we have the ZH bifurcation (\ref{eq:ZH}):

\begin{equation}
\mathcal{ZH}_3:~\mathcal{S}_3\cap\mathcal{H}_3\cap\{\vec{p} \cdot \vec{q} \ne 0\}
\end{equation}

\noindent
Here $\mathcal{S}_3\cap\mathcal{H}_3$ can be computed explicitly by
first solving $\mathcal{S}_3$ for $\sigma_3'$ and then substituting that solution into  $\widetilde{\mathcal{H}}_3$
and solving for $\sigma_2'$. This gives a manifold parameterized by $\sigma_1'$. However, the resulting expressions are quite complicated. This manifold must then be restricted by the conditions $\Delta_3<0$ and $\vec{p} \cdot \vec{q} \ne 0$, where $\vec{p}$ and $\vec{q}$ have already been computed for 3-neuron CTRNNs in (\ref{eq:qp3}).
For $\mat{W}_3$, these equations and restrictions can be solved exactly to obtain

\begin{equation*}
\begin{split}
\mathcal{ZH}_3:~(\sigma_1',\sigma_2',\sigma_3') = \Bigg\{&\left(\sigma_1',\frac{\alpha+\beta\epsilon}{\gamma+\delta},\frac{1}{4}\left(1-2\sigma_1'+\vert 1-6\sigma_1'\rvert \epsilon\right)\right),\\
&\left(\sigma_1',\frac{\alpha-\beta\epsilon}{\gamma-\delta},\frac{1}{4}\left(1-2\sigma_1'-\vert 1-6\sigma_1'\rvert \epsilon\right)\right)\Bigg\},\\
&\!\!\!\!\frac{1}{6}-\zeta-\eta < \sigma_1' < \frac{1}{6}-\zeta+\eta~~\textrm{or}~~\frac{5}{26} < \sigma_1' < \zeta
\end{split}
\end{equation*}

\noindent
where

\begin{equation*}
\begin{split}
\alpha \equiv&~2+\sigma_1'(74\sigma_1'-25)\\
\beta \equiv&~6-37\sigma_1'\lvert 1-6\sigma_1' \rvert\\
\gamma \equiv&~13+4\sigma_1'(117\sigma_1'-40)\\
\delta \equiv&~\sqrt{5-26\sigma_1'} \sqrt{37-234\sigma_1'} \lvert 1-6\sigma_1' \rvert\\
\epsilon \equiv&~\sqrt{\frac{5-26\sigma_1'}{37-234\sigma_1'}}\\
\zeta \equiv& \frac{\cos\left(\frac{1}{3}\arctan\left(2\sqrt{\frac{33}{37}}\right)\right)}{3\sqrt{37}}\\
\eta \equiv& \frac{\sin\left(\frac{1}{3}\arctan\left(2\sqrt{\frac{33}{37}}\right)\right)}{2\sqrt{111}}
\end{split}
\end{equation*}

\noindent
These restricted solutions produce the yellow curves of ZH bifurcations shown in Figure \ref{fig:CTRNN3}A.

Finally, all of the bifurcation manifolds derived above can be mapped from activation function 
derivative space $(\sigma_1',
\sigma_2',\sigma_3')$ to net input space $(\theta_1, \theta_2, \theta_3)$ using (\ref{eq:SigmaTheta}) and
(\ref{eq:DSigmaInverse}), once again taking care to maintain branch consistency. The result is shown in
Figure \ref{fig:CTRNN3}B. Here the combinatorial structure of $\mathcal{S}$ \cite{beer06,beer10} is becoming readily apparent.
The $\mathcal{BT}_3$, $\mathcal{GH}_3$ and $\mathcal{ZH}_3$
curves can also be used to determine the initial direction and shape of codimension-1 global bifurcation
surfaces for numerical continuation. In addition, points where multiple
codimension-2 bifurcation curves intersect indicate possible higher-codimension bifurcations.

\subsection{4-neuron CTRNNs}

The parameter space of a 4-neuron CTRNN is 20-dimensional and all of the bifurcations discussed in this
paper can occur generically. Rather than performing a complete analysis of a 4-neuron circuit, we illustrate
only the local codimension-2 bifurcation that is new in this case, namely the Hopf-Hopf bifurcation.

For a 4-neuron CTRNN, the pseudo-Hopf bifurcation condition in (\ref{eq:H}) can be written out explicitly
and solved to obtain an expression for $\sigma_4'$ with three branches that is too complicated
to reproduce here. By restricting these solutions to those with a pair of pure imaginary eigenvalues,
we obtain 3-dimensional Hopf manifolds in the 4-dimensional $\vec{\upsigma}'$ space.

Consider the example weight matrix

\begin{equation*}
\mat{W}_4 = 
\begin{pmatrix}
10 & -\frac{123}{10} & -\frac{47}{10} & -\frac{73}{5}\\[6pt]
8 & \frac{87}{10} & -\frac{78}{5} & -\frac{129}{10}\\[6pt]
\frac{41}{10} & \frac{3}{10} & -2 & \frac{25}{2}\\[6pt]
\frac{23}{5} & \frac{43}{10} & -\frac{71}{5} & 6\\
\end{pmatrix}
\end{equation*}

\begin{figure}[t]
\includegraphics[width=12cm]{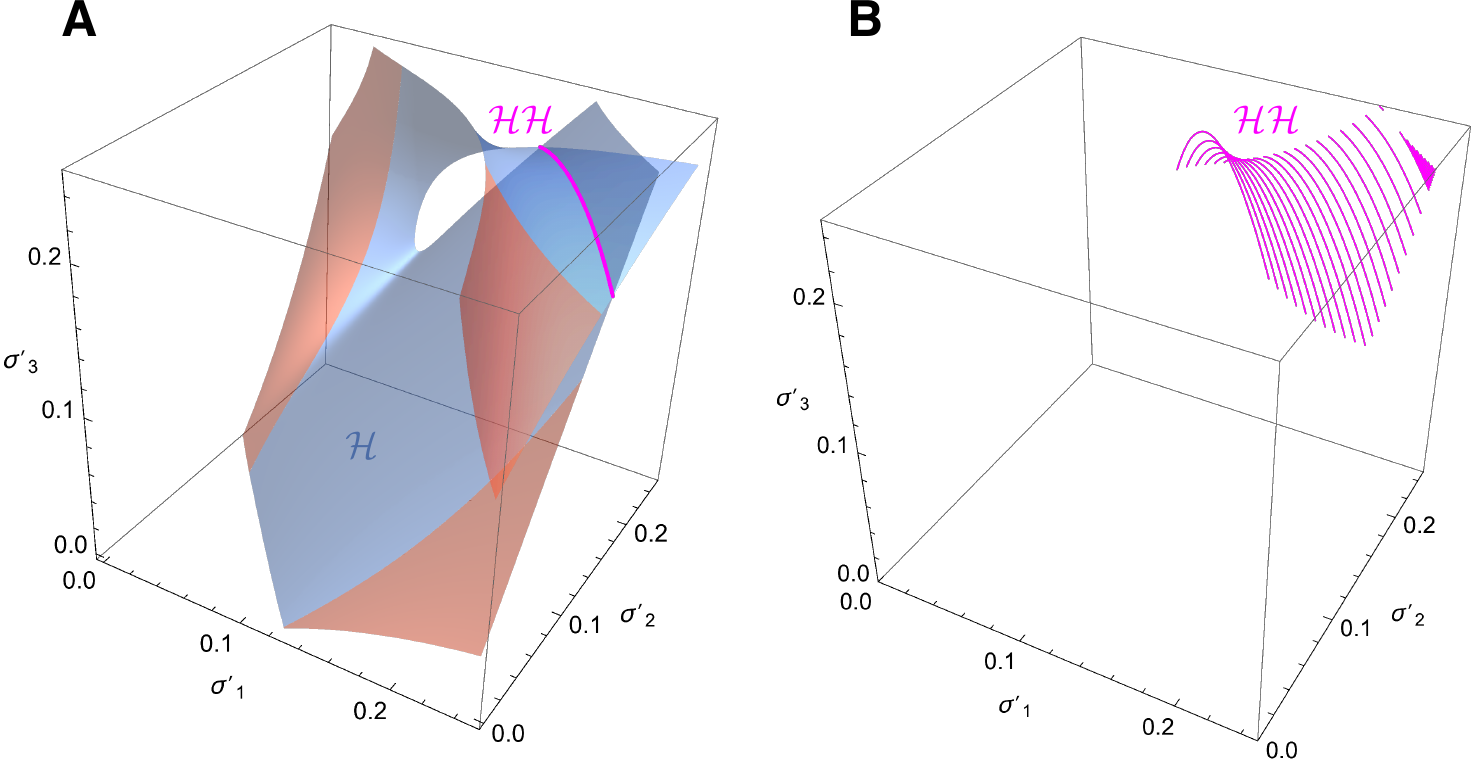}
\caption{A 3-dimensional projection of Hopf-Hopf bifurcations in the 4-neuron $\sigma$-CTRNN given by 
$\mat{W}_4$. (A) The 2D $\sigma_4'=1/8$ contour of the 3D pseudo-Hopf manifold $\widetilde{\mathcal{H}}_4$ 
embedded in 4D $\vec{\upsigma}'$ space. The blue portions of this surface represent the true Hopf bifurcation 
manifold $\mathcal{H}_4$, whereas the red portions correspond to neutral saddles. The magenta line
of self-intersection of $\mathcal{H}_4$ corresponds to a curve of Hopf-Hopf bifurcations along this contour. 
(B) The union of such curves for all $0<\sigma_4'\le1/4$ traces out the full 2D surface of $\mathcal{HH}_4$.}
\centering
\label{fig:CTRNN4}
\end{figure}

\noindent
The resulting $(\sigma_1',\sigma_2',\sigma_3')$ projection of the three branches of $\widetilde{\mathcal{H}}_4$ 
for a single $\sigma_4'$ contour of $\mat{W}_4$ are shown in Figure \ref{fig:CTRNN4}A, with the true
Hopf surface $\mathcal{H}_4$ indicated in blue and the portion of $\widetilde{\mathcal{H}}_4$ corresponding
to neutral saddles colored red.

In order to compute the 2-dimensional Hopf-Hopf manifold $\mathcal{HH}_4$ in $\mat{W}_4$, we must identify
the subset of $\mathcal{H}_4$ exhibiting a second pair of pure imaginary eigenvalues.
One way to proceed is as follows. 
First, as described above, we explicitly write out the pseudo-Hopf condition 
$\widetilde{\mathcal{H}}_4$ for $\mat{W}_4$ and solve it for $\sigma_4'$. We then apply the restrictions 
$\Delta_4>0,P>0,D>0$  involving the discriminant and two related polynomials of the characteristic equation of 
$\mat{A}_4$ \cite{rees}, where

\begin{equation*}
\begin{split}
P \equiv &~8\mu_2-3\tau^2\\
D \equiv &~64\det{\mat{A}_4}-16\mu_2^2+16\tau^2\mu_2-16\tau \mu_3-3\tau^4
\end{split}
\end{equation*}

\noindent
Together, $\widetilde{\mathcal{H}}_4$ and the above restrictions guarantee that two eigenvalues are of the form
$\pm\omega_1 i$ and the other two are of the form $\alpha\pm\omega_2 i$. Thus, the necessary condition for
$\mathcal{HH}_4$ is simply $\alpha = 0$, with additional nondegeneracy and transversality conditions required 
for sufficiency. Interestingly, for $\mat{W}_4$ Hopf-Hopf bifurcations correspond to transversal
self-intersections of $\mathcal{H}_4$ (e.g., magenta line in Figure \ref{fig:CTRNN4}A).
A 3-dimensional projection of the resulting 2-dimensional manifold $\mathcal{HH}_4$ is shown
in Figure \ref{fig:CTRNN4}B. As always, these manifolds can be mapped from $\vec{\upsigma}'$ space into
$\vec{\uptheta}'$ 
space using (\ref{eq:SigmaTheta}) and (\ref{eq:DSigmaInverse}), taking care to maintain branch consistency.

\section{Conclusion}

If we are to ever move beyond the study of isolated special cases in theoretical neuroscience,
we need to develop a more general theory of neural circuits over a given neural model.
Is such a thing even possible? What might such a theory look like? Where should we even begin? 
One idea is to select some particularly simple but nontrivial toy model neuron and work incrementally
from smaller to larger circuits, mapping out the parameter space structure of its dynamics as
completely as possible at each step. Such an endeavor can not only help to point the way toward
a more general theory of neural circuits, but may also be
independently useful to the extent that the selected model is utilized in computational neuroscience
and neural network applications.

This paper has taken another small step toward that goal.
Building upon previous work characterizing the local codimension-1 bifurcation structure of CTRNN
parameter space, we have investigated the local codimension-2 bifurcation structure of
CTRNNs. Specifically, we have (1) derived the necessary conditions for all generic
local codimension-2 bifurcations for general CTRNNs, (2) specialized these conditions to $\sigma$-CTRNNs
containing from one to four neurons, (3) illustrated in full detail the application of these conditions to 
example circuits, (4) derived closed-form expressions for these bifurcation manifolds where possible, and
(5) demonstrated how this analysis allows us to find and trace several global codimension-1 bifurcation manifolds that originate from these codimension-2 points.

Several directions for future work can be identified. First, all of the
above derivations should be repeated with the time constant parameters restored. Second, it would be 
useful to automate the symbolic derivation of the bifurcation conditions and, when possible, their
solutions for CTRNNs of particular sizes using a computer algebra system. Third, an attempt should
be made to develop simpler 
approximations for the various codimension-2 manifolds described here, as was done previously for the 
local codimension-1 bifurcation manifolds \cite{beer06}.
Fourth, it would be interesting to extend the analysis to even higher codimension bifurcations.
Codimension-2 bifurcations are themselves organized by codimension-3 bifurcations, and so on. 
In fact, we have seen some examples of apparent higher-codimension bifurcations in the specific circuits
examined above. This suggests identifying 
and analyzing the highest codimension bifurcations that a CTRNN of a given size
can exhibit and then tracing down to lower and lower codimension bifurcations as a strategy for
characterizing the overall structure of CTRNN parameter space. Finally, as progress on the CTRNN
toy model is made,
it will be important to incrementally extend the approach to more complicated neural models.

\clearpage
\printbibliography

@article{barak,
    author = "O. Barak",
    title = "Recurrent neural networks as versatile tools of neuroscience research",
    journal = "Current Opinion in Neurobiology",
    year = "2017",
    volume = "46",
    pages = "1--6"
}

@article{beer92,
    author = "R.D. Beer and J.C. Gallagher",
    title = "Evolving dynamical neural networks for adaptive behavior",
    journal ="Adaptive Behavior",
    year = "1992",
    volume = "1",
    pages = "91--122"
}

@article{beer95,
    author = "R.D. Beer",
    title = "On the dynamics of small continuous-time recurrent neural networks",
    journal = "Adaptive Behavior",
    year = "1995",
    volume = "3",
    pages = "471--511"
}

@article{beer06,
    author = "R.D. Beer",
    title = "Parameter space structure of continuous-time recurrent neural networks",
    journal = "Neural Computation",
    year = "2006",
    volume = "18",
    pages = "3009--3051"
}

@misc{beer10,
      author = "R.D. Beer and B. Daniels",
      title = "Saturation probabilities of continuous-time sigmoidal networks", 
      year = "2010",
      eprint = "1010.1714",
      archivePrefix = "arXiv",
      primaryClass = "q-bio.NC"
}

@article{blum,
    author = "E.K. Blum and X. Wang",
    title = "Stability of fixed points and periodic orbits and bifurcations in analog neural networks",
    journal = "Neural Networks",
    year = "1992",
    volume = "5",
    pages = "577--587"
}

@article{borisyuk,
    author = "R.M. Borisyuk and A.B. Kirillov",
    title = "Bifurcation analysis of a neural network model",
    journal = "Biological Cybernetics",
    year = "1992",
    volume = "66",
    pages = "319--325"
}

@article{cervantes-ojeda,
    author = "J. Cervantes-Ojeda and M. G\'{o}mez-Fuentes and R. Bernal-Jaquez",
    title = "Empirical analysis of bifurcations in the full weights space of a two-neuron DTRNN",
    journal = "Neurocomputing",
    year = "2017",
    volume = "237",
    pages = "362--374"}

@article{chow,
    author = "T.W.S. Chow and X.-D. Li",
    title = "Modeling of continuous time dynamical systems with input by recurrent neural networks",
    year = "2000",
    journal = "IEEE Transactions on Circuits and Systems—I: Fundamental Theory and Applications",
    volume = "47",
    pages = "575-—578"
}

@article{cohen,
    author = "M.A. Cohen and S. Grossberg",
    title = "Absolute stability of global pattern formation and parallel memory storage by competitive neural networks",
    year = "1983",
    journal = "IEEE Transactions on Systems, Man and Cybernetics",
    volume = "13",
    pages = "813--825"
}

@inbook{cowan,
    author = "J.D. Cowan and G.B. Ermentrout",
    title = "Some aspects of the eigenbehavior of neural nets",
    editor = "S.A. Levin",
    booktitle = "Studies in Mathematical Biology 1: Cellular Behavior and the Development of Pattern",
    publisher = "The Mathematial Association of America",
    year = "1978",
    pages = "67--117"
}

@article{das,
    author = "P.K. Dasa and W.C. Schieve and Z. Zheng",
    title = "Chaos in an effective four-neuron neural network",
    journal = "Physics Letters A",
    year = "1991",
    volume = "161",
    pages = "60--66"
}

@book{ermentrout,
    author = "G.B. Ermentrout and D.H. Terman",
    title = "Mathematical Foundations of Neuroscience",
    year = "2010",
    publisher = "Springer"
}

@article{fasoli,
    author = "D. Fasoli and A. Cattani and S. Panzeri",
    title = "The complexity of dynamics in small neural circuits",
    journal = "PLoS Computational Biology",
    year = "2016",
    volume = "12",
    pages = "e1004992"
}

@inbook{floreano,
    author = "S. Nolfi and J. Bongard and P. Husbands and D. Floreano",
    title = "Evolutionary robotics",
    editor = "B. Siciliano and O. Khatib",
    booktitle = "Springer Handbook of Robotics",
    publisher = "Springer",
    year = "2016",
    pages = "2030--2068"
}

@article{funahashi,
    author = "K.I. Funahashi and Y. Nakamura",
    title = "Approximation of dynamical systems by continuous time recurrent neural networks",
    year = "1993",
    journal = "Neural Networks",
    volume = "6",
    pages = "801-—806"
}

@article{gao,
    author = "P. Gao and S. Ganguli",
    title = "On simplicity and complexity in the brave new world of large-scale neuroscience",
    journal = "Current Opinion in Neurobiology",
    year = "2015",
    volume = "32",
    pages = "148--155"
}

@article{grossberg,
    author = "S. Grossberg",
    title = "On learning and energy-entropy dependence in recurrent and nonrecurrent signed
    networks",
    journal = "Journal of Statistical Physics",
    year = "1969",
    volume = "1",
    pages = "319--350"
}

@article{guckenheimer,
    author = "J. Guckenheimer and M. Myers and B. Sturmfels",
    title = "Computing Hopf bifurcations I",
    journal = "SIAM Journal on Numerical Analysis",
    year = "1997",
    volume = "34",
    pages = "1-21"
}

@article{haschke,
    author = "R. Haschke and J.J. Steil",
    title = "Input space bifurcation manifolds of recurrent neural networks",
    journal = "Neurocomputing",
    year = "2004",
    volume = "64C",
    pages = "25--38"
}

@article{hirsch,
    author = "M. Hirsch",
    title = "Convergent activation dynamics in continuous time networks",
    journal = "Neural Networks",
    year = "1989",
    volume = "2",
    pages = "331--349"
}

@article{hopfield,
    author = "J.J. Hopfield",
    title = "Neurons with graded response have collective computational properties like those of two-state neurons",
    journal = "Proceedings of the National Academy of Sciences",
    year = "1984",
    volume = "81",
    pages = "3088--3092"
}

@article{hopfieldtank,
    author = "J.J. Hopfield and D.W. Tank",
    title = "``Neural'' computation of decisions in optimization problems",
    journal = "Biological Cybernetics",
    year = "1985",
    volume = "52",
    pages = "141-152"
}

@book{hoppensteadt,
    author = "F.C. Hoppensteadt and E.M. Izhikevich",
    title = "Weakly Connected Neural Networks",
    year = "1997",
    publisher = "Springer"
}

@article{jaeger,
    author = "H. Jaeger and H. Haas",
    title = "Harnessing nonlinearity: predicting chaotic systems and saving energy in wireless communication",
    journal = "Science",
    year = "2004",
    volume = "304",
    pages = "78--80"
}

@article{izquierdo10,
    author = "E.J. Izquierdo and S.R. Lockery",
    title = "Evolution and analysis of minimal neural circuits for klinotaxis in \em{C. elegans}",
    journal = "Journal of Neuroscience",
    year = "2010",
    volume = "30",
    pages = "12908--12817"
}

@article{kimura,
    author = "M. Kimura and R. Nakano",
    title = "Learning dynamical systems by recurrent neural networks from orbits",
    year = "1998",
    journal = "Neural Networks",
    volume = "11",
    pages = "1589-—1599"
}

@book{kuznetsov,
    author = "Y.A. Kuznetsov",
    title = "Elements of Applied Bifurcation Theory",
    edition = "3",
    year = "2004",
    publisher = "Springer"
}

@article{maass,
    author = "W. Maass and T. Natschläger and H. Markram",
    title = "Real-time computing without stable states: a new framework for neural computation based on perturbations",
    journal = "Neural Computation",
    year = "2002",
    volume = "14",
    pages = "2531--2560"
    }

@article{mathayomchan,
    author = "B. Mathayomchan and R.D. Beer",
    title = "Center-crossing recurrent neural networks for the evolution of rhythmic behavior",
    journal = "Neural Computation",
    year = "2002",
    volume = "14",
    pages = "2043--2051"
}

@article{olivares,
    author = " E.O. Olivares and E.J. Izquierdo and R.D. Beer",
    title = "Potential role of a ventral nerve cord central pattern generator in forward and backward locomotion in \em{Caenorhabditis elegans}",
    journal = "Network Neuroscience",
    year = "2018",
    volume = "2",
    pages = "323-–343"
}

@article{pasemann,
    author = "F. Pasemann",
    title = "Complex dynamics and the structure of small neural networks",
    journal = "Network: Computation in Neural Systems",
    year = "2002",
    volume = "13",
    pages = "195--216"
}

@article{rees,
    author = "E.L. Rees",
    title = "Graphical discussion of the roots of a quartic equation",
    journal = "The American Mathematical Monthly",
    year = "1922",
    volume = "29",
    pages = "51--55"
}

@article{sompolinsky,
    author = "H. Sompolinsky and A. Crisanti",
    title = "Chaos in random neural networks",
    journal = "Physical Review Letters",
    year = "1988",
    volume = "61",
    pages = "259--262"}

@article{smithmiles,
    author = "K. Smith-Miles",
    title = "Neural networks for combinatorial optimization: A review of more than a decade of research",
    journal = "INFORMS Journal on Computing",
    year = "1999",
    volume = "11",
    pages = "15--34"
}

@article{sussillo,
    author = "D. Sussillo",
    title = "Neural circuits as computational dynamical systems",
    journal = "Current Opinion in Neurobiology",
    year = "2014",
    volume = "25",
    pages = "156–-163"
}

@article{tino,
    author = "P. Tin\v{o} and B.G. Horne and C.L. Giles",
    title = "Attractive periodic sets in discrete-time recurrent neural networks (with emphasis on fixed-point stability and bifurcations in two-neuron networks)",
    journal = "Neural Computation",
    year = "2001",
    volume = "13",
    pages = "1379--1414"
}

@book{wiggins,
    author = "S. Wiggins",
    title = "Introduction to Applied Nonlinear Dynamical Systems and Chaos",
    edition = "2",
    year = "2003",
    publisher = "Springer"
}

@article{wilson,
    author = "H.R. Wilson and J.D. Cowan",
    title = "Excitatory and inhibitory interactions in localized populations of model neurons",
    journal = "Biophysics Journal",
    year = "1972",
    volume = "12",
    pages = "1--24"
}

@article{yu,
    author = "Y. Yu and X. Si and C. Hu and J. Zhang",
    title = "A review of recurrent neural networks: LSTM cells and network architecture",
    journal = "Neural Computation",
    year = "2019",
    volume = "31",
    pages = "1235--1270"
}

@article{zhaojue,
    author = "Z. Zhaojue and W.C. Schieve and P.K. Das",
    title = "Two neuron dynamics and adiabatic elimination",
    journal = "Physica D",
    year = "1993",
    volume = "67",
    pages = "224--236"
}

\end{document}